\documentclass[draft]{agujournal2019}
\usepackage{url} %this package should fix any errors with URLs in refs.
\usepackage[inline]{trackchanges} %for better track changes. finalnew option will compile document with changes incorporated.
\usepackage{soul}
\draftfalse

\journalname{JGR: Planets}

\begin{document}

%%%%%%%%%%%%%%%%%%%%%%%%%%%%%%%%%%%%%%%%%%%%%%%
%  TITLE
%
% (A title should be specific, informative, and brief. Use
% abbreviations only if they are defined in the abstract. Titles that
% start with general keywords then specific terms are optimized in
% searches)
%
%%%%%%%%%%%%%%%%%%%%%%%%%%%%%%%%%%%%%%%%%%%%%%%

% Example: \title{This is a test title}

\title{Quantifying Surface Heterogeneity Across Asteroid (101955) Bennu using Candidate Site Remote Sensing Data}

%%%%%%%%%%%%%%%%%%%%%%%%%%%%%%%%%%%%%%%%%%%%%%%
%
%  AUTHORS AND AFFILIATIONS
%
%%%%%%%%%%%%%%%%%%%%%%%%%%%%%%%%%%%%%%%%%%%%%%%

% Authors are individuals who have significantly contributed to the
% research and preparation of the article. Group authors are allowed, if
% each author in the group is separately identified in an appendix.)

% List authors by first name or initial followed by last name and
% separated by commas. Use \affil{} to number affiliations, and
% \thanks{} for author notes.
% Additional author notes should be indicated with \thanks{} (for
% example, for current addresses).

% Example: \authors{A. B. Author\affil{1}\thanks{Current address, Antartica}, B. C. Author\affil{2,3}, and D. E.
% Author\affil{3,4}\thanks{Also funded by Monsanto.}}

\authors{E. C. Belhadfa\affil{1}, N. E. Bowles\affil{1}, K. A. Shirley \affil{1}, A. A. Simon \affil{2}, V. E. Hamilton \affil{3}, H. H. Kaplan \affil{2}}

% \affiliation{1}{First Affiliation}
% \affiliation{2}{Second Affiliation}
% \affiliation{3}{Third Affiliation}
% \affiliation{4}{Fourth Affiliation}

\affiliation{1}{Atmospheric, Oceanic and Planetary Physics, Clarendon Laboratory, University of Oxford, Oxford, United Kingdom}
\affiliation{2}{NASA Goddard Space Flight Centre, United States of America}
\affiliation{3}{Southwest Research Institute,  United States of America}
%(repeat as many times as is necessary)

% Corresponding author mailing address and e-mail address:

% (include name and email addresses of the corresponding author.  More
% than one corresponding author is allowed in this LaTeX file and for
% publication; but only one corresponding author is allowed in our
% editorial system.)

% Example: \correspondingauthor{First and Last Name}{email@address.edu}

\correspondingauthor{Emma-Catherine Belhadfa}{emma.belhadfa@physics.ox.ac.uk}

%%%%%%%%%%%%%%%%%%%%%%%%%%%%%%%%%%%%%%%%%%%%%%%
% KEY POINTS
%%%%%%%%%%%%%%%%%%%%%%%%%%%%%%%%%%%%%%%%%%%%%%%
%  List up to three key points (at least one is required)
%  Key Points summarize the main points and conclusions of the article
%  Each must be 140 characters or fewer with no special characters or punctuation and must be complete sentences

% Example:
% \begin{keypoints}
% \item	List up to three key points (at least one is required)
% \item	Key Points summarize the main points and conclusions of the article
% \item	Each must be 140 characters or fewer with no special characters or punctuation and must be complete sentences
% \end{keypoints}

\begin{keypoints}
\item OSIRIS-REx VNIR and TIR spectra of Bennu's four candidate sites show statistically significant spectral differences at 2--10\,m scales.
\item Bennu's regolith preserves spatially variable aqueous alteration at meter scales, reflecting a heterogeneous hydrothermal history.
\item Nightingale's spectral parameters span the full compositional range of all four sites, supporting its representativeness for sample return.
\end{keypoints}

%%%%%%%%%%%%%%%%%%%%%%%%%%%%%%%%%%%%%%%%%%%%%%%
%
%  ABSTRACT and PLAIN LANGUAGE SUMMARY
%
% A good Abstract will begin with a short description of the problem
% being addressed, briefly describe the new data or analyses, then
% briefly states the main conclusion(s) and how they are supported and
% uncertainties.

% The Plain Language Summary should be written for a broad audience,
% including journalists and the science-interested public, that will not have 
% a background in your field.
%
% A Plain Language Summary is required in GRL, JGR: Planets, JGR: Biogeosciences,
% JGR: Oceans, G-Cubed, Reviews of Geophysics, and JAMES.
% see http://sharingscience.agu.org/creating-plain-language-summary/)
%
%%%%%%%%%%%%%%%%%%%%%%%%%%%%%%%%%%%%%%%%%%%%%%%

%% \begin{abstract} starts the second page

\begin{abstract}
The OSIRIS-REx mission acquired spatially resolved (2-10 m spot sizes) visible--near infrared (VNIR) and thermal infrared (TIR) spectra across four candidate sampling sites on asteroid (101955) Bennu: Nightingale, Osprey, Sandpiper, and Kingfisher. To quantify heterogeneity across a small body ($\sim500$ m radius) like Bennu, we explore remotely observed spectral data to draw conclusions about the mineralogical composition and key physical processes that drive surface variability. We derive diagnostic band parameters from the OSIRIS-REx Visible and Infrared Spectrometer and the OSIRIS-REx Thermal Emission Spectrometer datasets to quantify compositional and physical variability across sites and assess their mineralogical context. The VNIR spectra exhibit similar overall reflectance shapes but systematic differences in spectral slopes and the 2.74 $\mu$m OH absorption. TIR emissivity spectra reveal modest but statistically significant shifts in the Christiansen Feature, silicate stretching, and bending band positions, indicating differences in silicate composition, hydration state, and Mg/Fe relative abundance. Principal component analysis separates each site into distinct clusters in multivariate band-parameter space, whereas K-means clustering identifies intra-site spectral sub-populations. Welch's Analysis of Variance and Hotelling’s $T^{2}$ tests confirm that band-parameter variations between sites are significant. These results reveal that Bennu's surface preserves measurable spectral heterogeneity at 2–10 m scales, with site-to-site variations in hydration indicators and silicate band positions. The spectral properties of Nightingale encompass the full range observed across all four sites, establishing a remote sensing baseline for contextualizing laboratory analyses of the returned sample within Bennu's broader composition diversity and alteration history.
\end{abstract}

\section*{Plain Language Summary}
NASA’s OSIRIS-REx spacecraft studied the surface of asteroid (101955) Bennu before collecting a sample in 2020. Four possible sample sites were mapped in detail using instruments that measure spectra that reveal what minerals are present and how the surface has been shaped over time.

In this study, we compare the spectra from the four candidate sample sites to determine how much variation can be captured on small spatial scales across Bennu. We find that each site has small, measurable differences in minerals and surface texture. Some sites show stronger signs of minerals that formed in the presence of water whereas others show features linked to different rock types or grain sizes. Using statistical tests, we show that the four sites are distinct from one another, and that the site ultimately chosen for sampling (Nightingale) covers the widest range of spectral properties.

Therefore, the returned sample captures the diversity present among these sites rather than representing just one compositional end-member. These findings will help scientists connect what OSIRIS-REx saw from orbit to what is measured directly in the laboratory, improving our understanding of how Bennu formed and evolved, and how water-rich materials may have moved through the early Solar System.

%%%%%%%%%%%%%%%%%%%%%%%%%%%%%%%%%%%%%%%%%%%%%%%
%
%  BODY TEXT
%
%%%%%%%%%%%%%%%%%%%%%%%%%%%%%%%%%%%%%%%%%%%%%%%

%%% Suggested section heads:
% \section{Introduction}
%
% The main text should start with an introduction. Except for short
% manuscripts (such as comments and replies), the text should be divided
% into sections, each with its own heading.

% Headings should be sentence fragments and do not begin with a
% lowercase letter or number. Examples of good headings are:

% \section{Materials and Methods}
% Here is text on Materials and Methods.
%
% \subsection{A descriptive heading about methods}
% More about Methods.
%
% \section{Data} (Or section title might be a descriptive heading about data)
%
% \section{Results} (Or section title might be a descriptive heading about the
% results)
%
% \section{Conclusions}
%%%%%%% INTRO %%%%%%%%%%%%%%
\section{Introduction}
NASA's Origins, Spectral Interpretation, Resource Identification, and Security – Regolith Explorer (OSIRIS-REx) mission provided the opportunity to investigate the physical, chemical, and mineralogical properties of a primitive near-Earth asteroid. Asteroid (101955) Bennu, the target of OSIRIS-REx, is a low-albedo, carbonaceous (B-type) asteroid with a diameter of $\sim$500~m and a rubble-pile structure \cite{Lauretta2019TheBennu, Dellagiustina2020VariationsBennu}. Primitive asteroids like Bennu have been found to contain hydrated clay minerals and organics that may have transported the building blocks for life to Earth \cite{Oba2022IdentifyingMeteorites}. Hence, the study of Bennu, through a detailed remote sensing campaign and the subsequent sample collection, yields insight into some of the most primitive material in our solar system, informing how the planets -- and life itself -- formed. 

During the mission’s Reconnaissance phase, multiple instruments, including the OSIRIS-REx Visible and Infrared Spectrometer (OVIRS) and the OSIRIS-REx Thermal Emission Spectrometer (OTES), acquired spatially resolved (2-10 m spot size) spectral datasets across Bennu’s surface \cite{Reuter2018TheBennu, Christensen2018TheInstrument}. This spectral data enables direct investigation of surface heterogeneity through diagnostic absorption features of minerals spanning the visible-near infrared (VNIR) to the mid-thermal infrared (TIR) range, informing our understanding of Bennu's composition and geological evolution. 

OVIRS and OTES captured core mineralogical abundances on Bennu. Magnetite was first detected by OTES \cite{Hamilton2019EvidenceBennu}, and then by OCAMS \cite{DellaGiustina2019PropertiesAnalysis} and OVIRS \cite{Kaplan2020VisiblenearOSIRIS-REx}. Fundamental absorption features of phyllosilicates were identified by OTES \cite{Hamilton2019EvidenceBennu} and then supported by OVIRS \cite{Zou2021PhotometryOSIRIS-REx} whereas bright carbonate veins were detected by OVIRS and OTES \cite{Kaplan2020BrightHistory}, both of which are indicative of widespread aqueous alteration in Bennu's history. Shifts in the Restrahlen bands, specifically the Si--O stretching minimum, is discussed in \citeA{Hamilton2021EvidenceSpectroscopy} and attributed to particle size effects based on multiple band positions.  

Surface heterogeneity on small bodies is driven by regolith evolution, space weathering, and compositional diversity at scales relevant to sample return \cite{Kaplan2020VisiblenearOSIRIS-REx, Dellagiustina2020VariationsBennu, Hamilton2021EvidenceSpectroscopy}. Subtle differences in dynamic surfaces processes and volatile transport can dramatically impact the composition and state of a sample. Samples provide the opportunity to study pristine asteroidal material in laboratory contexts at a resolution and accuracy greater than that provided by remote sensing alone. However, our understanding of a sample's representativeness is limited by our study of the context from which it was collected, motivating the need for in depth evaluation of surface heterogeneity across Bennu. Variations in spectral band parameters, such as absorption depth, band centre position, and spectral slope, reflect differences in mineralogy, grain size, and surface processing (e.g., impact gardening, thermal fracturing, or volatile mobility). On Bennu, the distribution of hydrated phyllosilicates, carbonates, and oxides such as magnetite has been well documented on global scales \cite{Hamilton2019EvidenceBennu, Kaplan2020BrightHistory, Kaplan2020VisiblenearOSIRIS-REx, Dellagiustina2020VariationsBennu}, yet the degree of variability across smaller spatial scales remains less constrained. Furthering our understanding of heterogeneity across these high-resolution remote sensing datasets will provide additional context needed to evaluate the representativeness of OSIRIS-REx's returned sample.

In this work, we quantify the spatial variability of Bennu’s surface composition by analyzing spectral band parameters derived from OVIRS reflectance spectra (0.4–4.3~$\mu$m) and OTES emissivity spectra (1750–100~cm$^{-1}$) \cite{Reuter2018TheBennu, Christensen2018TheInstrument}. Previous work by \citeA{Barucci2020OSIRIS-RExStatistics} applied a G-mode multivariate statistical approach to the global OVIRS dataset from the Equatorial Station~3 phase, identifying five spectral clusters at a $2\sigma$ confidence level and none at $3\sigma$. Their analysis suggested that Bennu’s surface is largely homogeneous at the spectral and spatial resolution of those data ($\sim$20--30~m per spot), with subtle slope variations concentrated around the equatorial ridge and large boulders such as Roc Saxum (Figure \ref{fig:samplesitesmap}). Importantly, no statistically significant difference was observed in the depth of the 2.74~$\mu$m hydration band, indicating a uniform abundance of hydrated phyllosilicates across the surface. This work builds upon \citeA{Barucci2020OSIRIS-RExStatistics} and other previous global-scale studies \cite{Hamilton2019EvidenceBennu, Simon2020WeakSpectrometer, Simon2020WidespreadBennu, Dellagiustina2020VariationsBennu} that explored other instrument datasets or resolutions. Here, we focus on site-specific differences by applying both quantitative spectral comparisons and statistical analyses across the four Reconnaissance sites (Nightingale, Osprey, Kingfisher, and Sandpiper). Our objectives are to:  
\begin{enumerate}
\label{objectives}
    \item Characterize the degree of heterogeneity in diagnostic absorption features across sampling sites across a broadband dataset in both VNIR and TIR regimes;
    \item Explore correlations between band parameters and potential compositional or physical drivers, such as phyllosilicate abundance, space weathering trends, or particle size effects;
    \item Assess the implications of spectral heterogeneity at small spatial scales for the interpretation of returned samples in the context of Bennu’s surface processes and geological history.
\end{enumerate}

By systematically comparing site-resolved spectral parameters, this study provides new discussions and constraints on metre-scale surface heterogeneity on Bennu and informs the representativeness of the returned sample within the context of the candidate sites.

%%%%%% METHODS %%%%%%%%%%

\section{Methods}
This section describes the datasets, calibration, preprocessing, band parameter definitions, and statistical framework used to quantify inter-site variability across Bennu’s four Reconnaissance candidate sites.

\subsection{Data}
\begin{figure}
    \centering
    \includegraphics[width=1\linewidth]{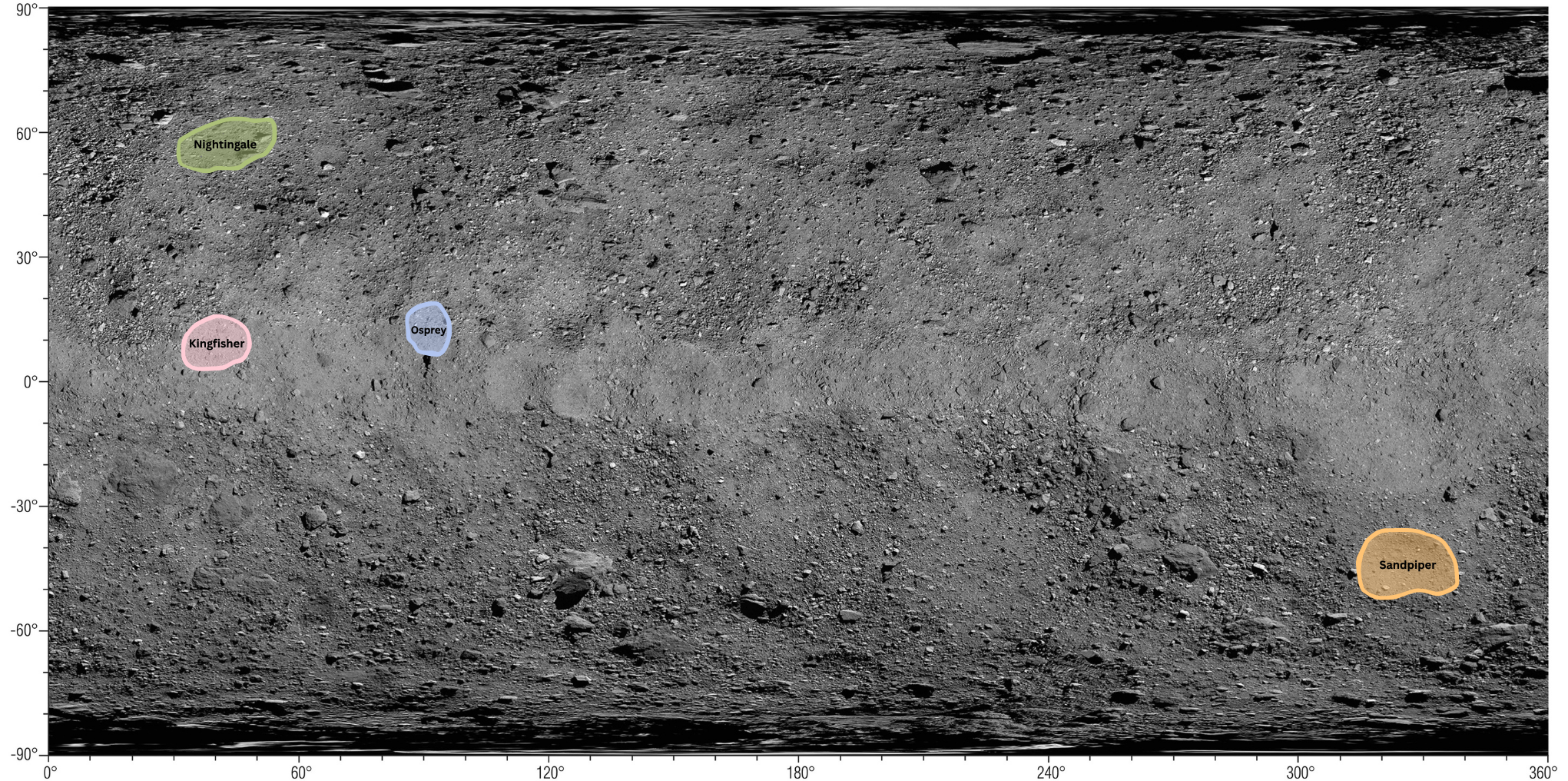}
    \caption{Locations of the four candidate sites across a 2D map projection of Bennu's surface captured by the OSIRIS-REx Camera Suite (OCAMS) \cite{Bennett2021ABennu}. All sample sites are outlined based on the footprint of data used in this work. Nightingale (Latitude: 56\textdegree, Longitude: 43\textdegree) is located in Hokioi crater (10 m radius) near Bennu’s north pole. Osprey (Latitude: 11\textdegree, Longitude: 88\textdegree) is set in a small crater with a 10 m radius. Kingfisher (Latitude: 11\textdegree, Longitude: 56\textdegree) is located in a small crater (10 m radius) and surrounded by boulders, but the site itself is free of large obstructions. Sandpiper (Latitude: -47\textdegree, Longitude: 322\textdegree) is located in Bennu’s southern hemisphere, on the southeast floor of a large crater with a radius of 31.5 m.}
    \label{fig:samplesitesmap}
\end{figure}

We use OSIRIS-REx Reconnaissance-phase observations from OVIRS and OTES instruments \cite{Reuter2018TheBennu, Christensen2018TheInstrument}. The four candidate sampling sites (Nightingale, Osprey, Kingfisher, and Sandpiper (Figure \ref{fig:samplesitesmap})) were observed during three dedicated campaigns (RECON A–C). Table~\ref{tab:data_overview} lists the visit dates of the data used in this work and the number of spectra retained per site after quality control (QC).

\begin{table}
\centering
\caption{Reconnaissance observations used in this study. Dates correspond to site visits; counts are the number of spectra retained for each instrument.}
\label{tab:data_overview}
\begin{tabular}{lllcc}
\hline
\textbf{Site} & \textbf{Phase(s)} & \textbf{Date(s)} & \textbf{$N_{\mathrm{OVIRS}}$} & \textbf{$N_{\mathrm{OTES}}$} \\
\hline
Nightingale & RECON A,B & 2019-10-26; 2020-01-22 & \textit{11972} & \textit{4387} \\
Osprey      & RECON A,B & 2019-10-12; 2020-02-11 & \textit{9251} & \textit{7182} \\
Sandpiper   & RECON A     & 2019-10-05                           & \textit{3260} & \textit{2840} \\
Kingfisher  & RECON A     & 2019-10-19                           & \textit{4809} & \textit{2236} \\
\hline
\end{tabular}
\end{table}

Calibrated spectra for both OVIRS and OTES were collected from  the archival Planetary Database System (PDS). For each visit date in Table~\ref{tab:data_overview}, Level-3 spectra were processed according to their metadata files, filtered for flagged off-nominal measurements, and then averaged spatially to represent each of the four sites. Individual spectra were used to identify and remove any measurements containing over-saturation or instrument artifacts \cite{Simon2021DerivationCalibration}. OVIRS spectra were normalized at 0.55~$\mu$m, which reduces absolute radiometric uncertainty from about 5~\% to much smaller channel to channel relative uncertainty of $\leq$ 1\% \cite{Simon2021DerivationCalibration}. A full description of  data processing pipelines used for this study can be found in the Supplementary Materials. Figure~\ref{fig:site_spectra} shows the site-mean spectra generated from the processed sets.

\begin{figure}
    \centering
    \includegraphics[width=1\linewidth]{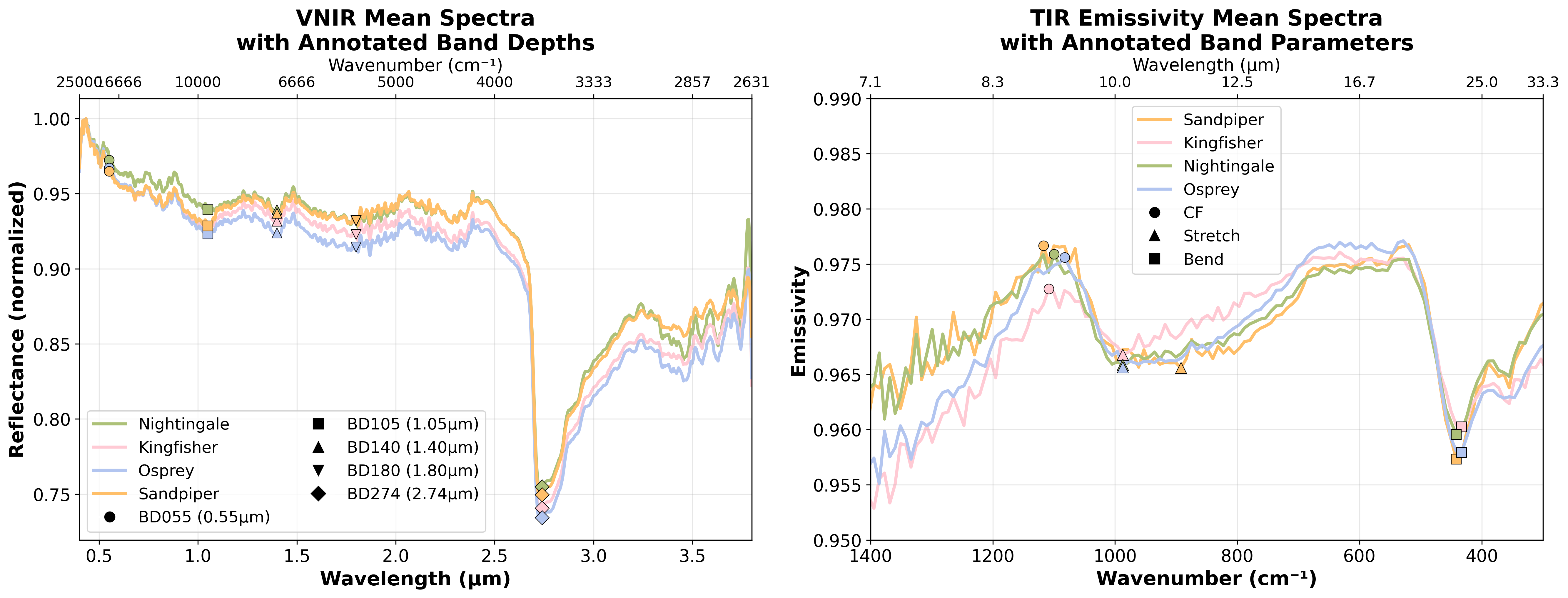}
    \caption{VNIR reflectance spectra collected by OVIRS and TIR emissivity spectra collected by OTES, averaged for each candidate site. Annotated with band parameters described in Section \ref{sec:bandparams}. In VNIR, the parameters are the band depths at 0.55, 1.05, 1.4, 1.8, and 2.74~$\mu$m and the 0.5–1.5~$\mu$m slope. In TIR, the parameters are the Christiansen Feature (CF) position, spectral slope, and Reststrahlen Band (Si-O) bend and stretch positions.}
    \label{fig:site_spectra}
\end{figure}

\subsection{Qualitative spectral context}
Before statistical modelling, site-mean spectra (Figure~\ref{fig:site_spectra}) were inspected to guide parameter selection. In VNIR \cite{Simon2020WidespreadBennu, Simon2020WeakSpectrometer, Praet2023EvaluatingSite}, a reproducible doublet near 1.3–1.5~$\mu$m is present at varying contrast; Sandpiper and Nightingale tend to cluster in 1.5–2.4~$\mu$m continuum shape. The 2.7~$\mu$m OH band varies in depth and/or width among sites, and the 3.3–3.5~$\mu$m region shows site-dependent patterns.

In TIR \cite{Hamilton2021EvidenceSpectroscopy}, spectral slopes through the partial transparency region above the CF span nearly two orders of magnitude across sites: Nightingale is the steepest ($-2.11 \times 10^{-4}$cm$^{1}$) and Kingfisher intermediate ($-1.40 \times 10^{-4}$cm$^{1}$), while Osprey ($+4.52 \times 10^{-5}$cm$^{1}$) and Sandpiper ($-4.91 \times 10^{-6}$cm$^{1}$) are flatter through this region. The CF position and amplitude vary modestly across sites, whereas the silicate-stretch minimum shifts by several tens of cm$^{-1}$. Past ${\sim}700$~cm$^{-1}$, a site-dependent broad Si--O bending feature around ${\sim}450$~cm$^{-1}$ contributes to overall low-wavenumber variability.

\subsection{Band Parameters}
\label{sec:bandparams}

To enable inter-site comparison, we determine band parameters that capture diagnostic absorptions and continuum behaviour in both VNIR and TIR. Band depths across three-channel windows, spectral slopes, and their associated uncertainties are calculated for each individual spectrum using the Spindex IDL system \citeA{Kaplan2020VisiblenearOSIRIS-REx}; site-mean parameters and propagated uncertainties ($\sqrt{\sum\sigma_i^{2}}/N$) are derived from the full per-spectrum distributions, with within-site standard deviations reported separately to characterise spectral heterogeneity (Figure~\ref{fig:violin}).

We initially considered the full weak-feature set of \citeA{Simon2020WeakSpectrometer} for OVIRS --- band depths at 0.55, 1.05, 1.4, 1.8, and 2.3~$\mu$m and the 0.5--1.5~$\mu$m continuum slope --- alongside the 2.74~$\mu$m OH absorption, and four OTES parameters (CF position, pre-CF spectral slope, Reststrahlen stretching minimum, and bending minimum). Following visual inspection, we eliminated the 2.3~$\mu$m feature (Mg/Fe phyllosilicates) as it was not consistently detectable above the noise level across all four sites at Reconnaissance-phase spatial resolution. The remaining parameters (Table~\ref{tab:bandparam_summary}) were selected on three grounds: (i) \textit{diagnostic value}: each parameter is sensitive to a distinct compositional or physical surface process, spanning hydration state, ferrous mineralogy, bulk silicate composition, and particle size; (ii) \textit{breadth}: together they address the objectives outlined in Section~\ref{objectives} across both VNIR and TIR regimes; and (iii) \textit{comparability}: all parameters are established in related studies of Bennu and analogous bodies \citeA{Viviano-Beck2014RevisedMars, Simon2020WeakSpectrometer, Bates2020LinkingReturn, Hamilton2021EvidenceSpectroscopy, Xie2022Thermal-InfraredAnalysis}, enabling direct comparison with results from other planetary bodies and materials.

However, important limitations apply to their interpretation. Band depth is not a uniquely interpretable parameter: observation and illumination geometry, particle size, and albedo all contribute to the measured values. Each parameter also covers a range of possible mineralogical and physical interpretations (Table~\ref{tab:bandparam_summary}), meaning our ability to isolate exact drivers of heterogeneity is inherently constrained in remote sensing datasets. This work aims to comment and quantify on the diversity and heterogeneity across a small-spatial scale dataset, rather than make causal determinations of spectral differences. 

\begin{sidewaystable}
\centering
\caption{Summary of Final Selected Band Parameters, Spectral Position/Ranges, and Interpretations}
\label{tab:bandparam_summary}
\resizebox{\textwidth}{!}{%
\begin{tabular}{lllll}
\hline
\textbf{Instrument} &
\textbf{Parameter} &
\textbf{Spectral Position/Range} &
\textbf{Mineralogical Interpretation} &
\textbf{Physical / Process Interpretation} \\
\hline
\multicolumn{5}{l}{\textit{\textbf{VNIR Reflectance Spectra}}} \\
\hline
OVIRS & BD$_{055}$ & 0.55 $\mu$m &
Magnetite / Fe$^{2+}$/Fe$^{3+}$ transition band &
Correlates with surface albedo and is sensitive to fine–grained coatings \\
OVIRS & BD$_{105}$ & 1.05 $\mu$m &
Ferrous silicates, Fe-carbonates, Fe-oxides, and some phyllosilicates &
Affected by grain size, porosity, and surface roughness \\
OVIRS & BD$_{140}$ & 1.4 $\mu$m &
OH hydration-sensitive band &
Proxy for bound water and influenced by space-weathering dehydration \\
OVIRS & BD$_{180}$ & 1.8 $\mu$m &
Combination OH/H$_2$O band in hydrated phases &
Sensitive to degree of aqueous alteration and mineral structure \\
OVIRS & BD$_{274}$ & 2.74 $\mu$m &
Structural OH in phyllosilicates (found CI/CM-like analogues) &
Primary indicator of hydration state and secondarily sensitive to thermal history \\
OVIRS & Slope$_{0.5-1.5}$ & 0.5–1.5 $\mu$m &
Broadband continuum tied to opaque mineral phases and alteration &
Proxy for space weathering, grain size, and micro-roughness \\
\hline
\multicolumn{5}{l}{\textit{\textbf{TIR Emission Spectra}}} \\
\hline
OTES & CF Position &
$\sim$1000–1200 cm$^{-1}$ &
Bulk silicate composition, shifted by Mg/Fe ratio or polymerization &
Shifted by both physical and compositional drivers, including grain size and hydration \\
OTES & TIR Slope &
1400 cm$^{-1}$ to CF &
Not mineralogically specific &
Linked to particle size distribution, where steeper slopes indicate finer grains \\
OTES & Stretch Minimum Position &
1100–800 cm$^{-1}$ &
Si–O stretching in phyllosilicates and anhydrous silicates &
Tracks OTES spectral types (Type 1–3) and fine particulate presence\\
OTES & Bend Minimum Position &
500–400 cm$^{-1}$ &
Si–O bending mode, linked to phyllosilicate content &
Correlates with alteration grade and is sensitive to mineral structural changes \\
\hline
\end{tabular}%
}
\begin{tablenotes}
\small
\textbf{Note:} Mineralogical and physical interpretations follow the frameworks established in \citeA{Hamilton2019EvidenceBennu, Simon2020WidespreadBennu,Hanna2020DistinguishingSpectroscopy,Hamilton2021EvidenceSpectroscopy, Clark2023OverviewBennu}.
\end{tablenotes}
\end{sidewaystable}

\subsection{Statistical Models}
\label{sec:stats}
Using a multi-model approach, we quantify heterogeneity across sites and within sites, identifying driving factors using variations in the defined band parameters. Custom Python scripts were developed in tandem with existing SciPy libraries. The code is available on the first author's GitHub.

The four sites differ substantially in sample size 
(Table~\ref{tab:data_overview}): OVIRS ranges from 3,260 (Sandpiper) to 11,972 (Nightingale), and OTES from 2,236 
(Kingfisher) to 7,182 (Osprey). This class imbalance, combined with the large absolute sample sizes, has implications for the sensitivity and interpretation of each statistical method described below, which are assessed in Section~\ref{sec:valid}. Additionally, spectra sampled within a single small site are spatially proximate and may not be fully independent, which is a recognized limitation of the frequentist framework applied here.

\textit{Principal Component Analysis (PCA)}: PCA is applied to the defined parameters to obtain low-dimensional embeddings and biplots, variance capture, and heuristics following \citeA{Greenacre2022PrincipalAnalysis, Xie2022Thermal-InfraredAnalysis}, allowing for the simplification of a complex, multi-dimensional dataset into geologically interpretable results. PERMANOVA \cite{Anderson2001AVariance} and PERMDISP \cite{Anderson2006Distance-basedDispersions} formally assess multivariate site separation using 999 permutations across the full suite of band parameters to distinguish between differences in group centroids and differences in within-site dispersion.

\textit{K-Means Clustering}: Unsupervised K-means clustering is applied to the PCA feature space to identify natural spectral groupings, both across all sites and within individual sites. For each dataset, solutions with k = 3 clusters were explored \cite{Macqueen1967SomeObservations, Lloyd1982LeastPCM}. 

\textit{Analysis of Variance (ANOVA)}: Welch's one-way ANOVA \cite{Welch1951OnApproach}, which relaxes the equal-variance assumption of traditional ANOVA, tests whether band-parameter means differ by site. Post-hoc pairwise inference uses Hotelling's \textit{T$^2$} test, which does not assume equal variances or equal sample sizes. 

\textit{Supporting inference tools}: To evaluate the significance of our results, we use Welch's $t$-tests (which assumes unequal variance) for targeted pairwise comparisons in small-$n$ contexts \cite{DeWinter2013UsingSizes}, F-statistics for model-level effect testing \cite{Fisher1925StatisticalWorkers}, and conventional $p$-value thresholds ($\alpha=0.01$) with interpretation per \citeA{Schervish1996PNot, Vidgen2016P-values:Misused} supplemented by Cohen's d metrics \cite{Cohen1988StatisticalScience}. 

\section{Results}

Spectral variability is quantified in this section using the multi-model approach previously outlined. 
\subsection{Spectral Band Parameter Comparisons}
\begin{figure}
    \centering
    \includegraphics[width=1\linewidth]{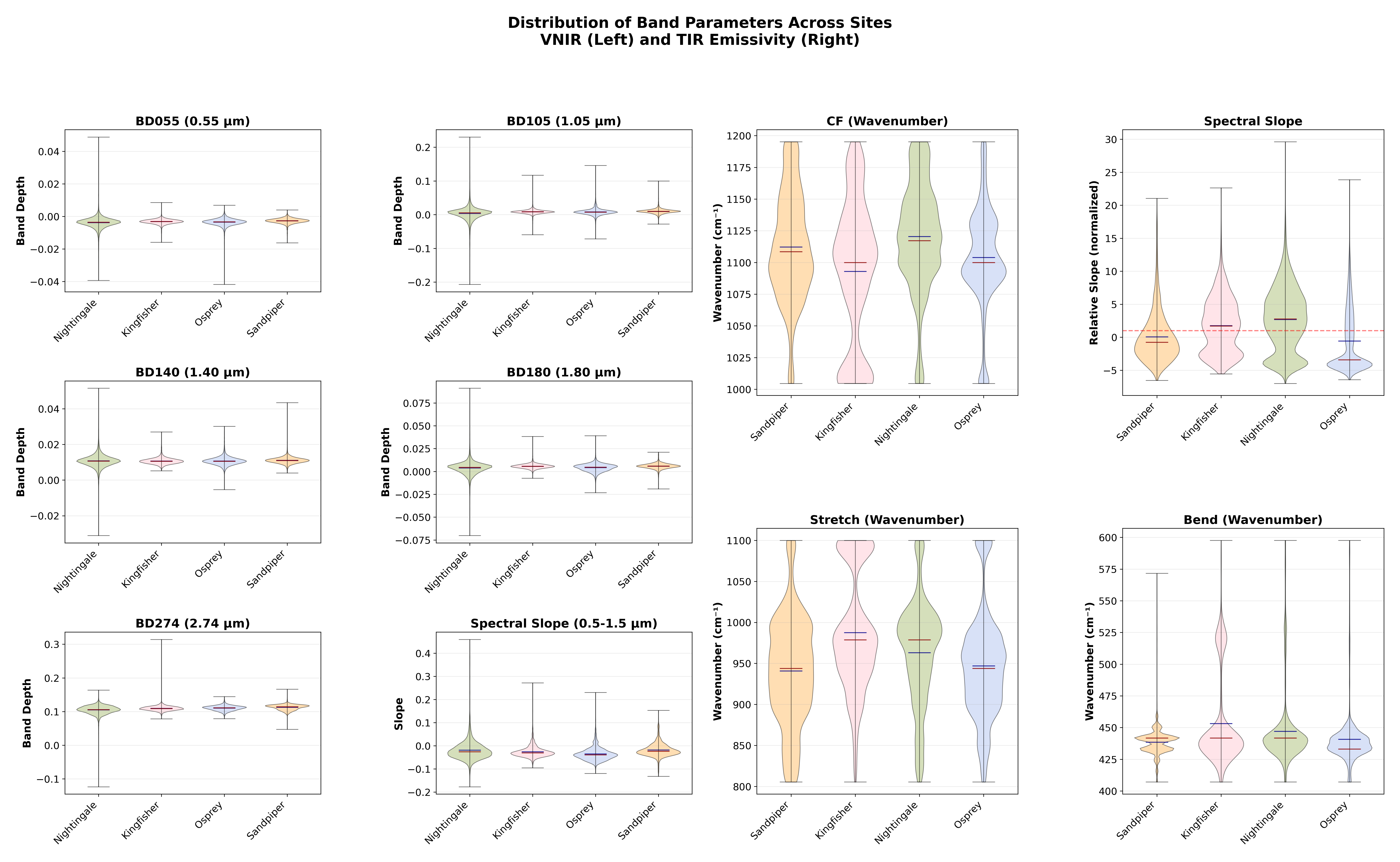}
    \caption{Violin plots of the selected band parameters for the OVIRS (left) and OTES (right) data, separated by site. The violin plots describe the probability density of the band parameters, with the mean (dark blue line) and median (dark red line) annotated. The violin represents the middle 50\% of the data, with its left edge being the first quartile (\(Q_{1}\)) and the right edge being the third quartile (\(Q_{3}\)). Whiskers extend from the box to the lowest and highest values.}
    \label{fig:violin}
\end{figure}

Figure 3 outlines the spread of the band parameter values across the four sites for all processed spectra. The violin plots describe the probability density, with the means and medians indicated. These distributions serve as a foundation to proceed with later quantification of separability and significance.

\begin{sidewaystable}
\caption{Mean and Standard Deviation of Band Parameters Across Candidate Sites}
\label{tab:band_param_statistics}x
\begin{tabular}{llcccc}
\hline
\textbf{Instrument} & \textbf{Parameter} & \textbf{Nightingale} & \textbf{Kingfisher} & \textbf{Osprey} & \textbf{Sandpiper} \\
\hline
\multicolumn{6}{l}{\textbf{OVIRS (VNIR)}} \\
& BD$_{0.55}$ & $-0.0038 \pm 3.5\times10^{-7}$ & $-0.0032 \pm 2.5\times10^{-7}$ & $-0.0035 \pm 3.0\times10^{-7}$ & $-0.0027 \pm 2.8\times10^{-7}$ \\
& BD$_{1.05}$ & $0.0037 \pm 1.1\times10^{-6}$ & $0.0085 \pm 4.4\times10^{-7}$ & $0.0082 \pm 4.9\times10^{-7}$ & $0.0099 \pm 7.1\times10^{-7}$ \\
& BD$_{1.40}$ & $0.0107 \pm 4.5\times10^{-7}$ & $0.0107 \pm 3.5\times10^{-7}$ & $0.0107 \pm 3.0\times10^{-7}$ & $0.0111 \pm 4.9\times10^{-7}$ \\
& BD$_{1.80}$ & $0.0039 \pm 1.3\times10^{-6}$ & $0.0057 \pm 3.6\times10^{-7}$ & $0.0042 \pm 3.4\times10^{-7}$ & $0.0059 \pm 4.8\times10^{-7}$ \\
& BD$_{2.74}$ & $0.1052 \pm 6.9\times10^{-6}$ & $0.1095 \pm 1.0\times10^{-5}$ & $0.1104 \pm 6.9\times10^{-6}$ & $0.1127 \pm 1.1\times10^{-5}$ \\
& Spectral Slope (1/$\mu$m) & $-1.894\times10^{-2} \pm 2.1\times10^{-5}$ & $-2.562\times10^{-2} \pm 2.0\times10^{-5}$ & $-3.545\times10^{-2} \pm 1.6\times10^{-5}$ & $-1.834\times10^{-2} \pm 2.5\times10^{-5}$ \\
\hline
\hline
\multicolumn{6}{l}{\textbf{OTES (TIR)}} \\
& CF Position (cm$^{-1}$) & $1121 \pm 43$ & $1093 \pm 56$ & $1104 \pm 41$ & $1112 \pm 43$ \\
& Spectral Slope (1/cm$^{-1}$) & $-2.099\times10^{-4} \pm 1.5\times10^{-6}$ & $-1.380\times10^{-4} \pm 2.0\times10^{-6}$ & $+4.302\times10^{-5} \pm 9.9\times10^{-7}$ & $-5.683\times10^{-6} \pm 1.8\times10^{-6}$ \\\\
& Restrahlen Stretch (cm$^{-1}$) & $963 \pm 65$ & $987 \pm 72$ & $947 \pm 65$ & $941 \pm 67$ \\
& Restrahlen Bend (cm$^{-1}$) & $447 \pm 29$ & $453 \pm 35$ & $441 \pm 24$ & $439 \pm 11$ \\
\hline
\end{tabular}
\end{sidewaystable}

\subsubsection{Reflectance (OVIRS)}
OVIRS reflectance spectra from the four candidate sites on Bennu exhibit broadly similar shapes but with notable differences in specific spectral features, shown in Figure \ref{fig:site_spectra}. All sites show a primary reflectance peak in the blue-green wavelengths ($\sim0.44~\mu$m) after normalization, indicating a generally consistent overall spectral shape. However, the depths of key absorption bands vary by site (Figure \ref{fig:violin}, left). For instance, the band depth at $\sim1.05~\mu$m (indicative of possible ferrous minerals) is deepest at Sandpiper ($0.0099 \pm 7.1\times10^{-7}$) and comparatively shallower at Nightingale ($0.0037 \pm 1.1\times10^{-6}$). A subtle $\sim0.55~\mu$m feature is present at all sites, but Nightingale shows slightly higher depths for this band than the other sites. 

In the longer wavelengths, Sandpiper’s spectra display the strongest $2.74~\mu$m hydration band, with a band depth roughly $\sim5\%$ greater than the site with the weakest hydration signal (Nightingale). Meanwhile, the spectral slope from $0.5$--$1.5~\mu$m varies: Osprey’s reflectance decreases most steeply with wavelength (a ``redder'' spectrum), whereas Sandpiper and Nightingale’s slope is more neutral. 

The small site-mean uncertainties in Table~\ref{tab:bandparam_summary} reflect the large number of spectra averaged per site. \citeA{Simon2021DerivationCalibration} demonstrated that the final OVIRS calibration enables reliable detection of band depths below1\%, and averaging over $N$ spectra reduces the standard error of the mean by $1/\sqrt{N}$, yielding site-mean uncertainties of order $10^{-6}$--$10^{-7}$.

\subsubsection{Thermal Emission (OTES)}
In the OTES emissivity data (Figure~\ref{fig:violin}, right), each site shows distinguishing characteristics aligned with compositional and/or physical differences. The CF position varies across sites from $1093 \pm 56$~cm$^{-1}$ (Kingfisher) to $1121 \pm 43$~cm$^{-1}$ (Nightingale), a site-to-site spread of $\sim$28~cm$^{-1}$ (Table~\ref{tab:band_param_statistics}). The Si-O stretching minimum also differs by site: Kingfisher sits highest at $988 \pm 72$~cm$^{-1}$, followed by Nightingale ($963 \pm 65$~cm$^{-1}$), Osprey ($947 \pm 65$~cm$^{-1}$), and Sandpiper lowest at $941 \pm 67$~cm$^{-1}$; Osprey and Sandpiper thus exhibit stretch minima shifted $\sim$41--47~cm$^{-1}$ lower than Kingfisher, consistent with more pronounced Reststrahlen band contrast at those sites. The Si-O bending mode minimum follows a similar ordering: Osprey ($441 \pm 24$~cm$^{-1}$) and Sandpiper ($439 \pm 11$~cm$^{-1}$) sit closest to the global average of 440~cm$^{-1}$ \cite{Hamilton2021EvidenceSpectroscopy}, while Kingfisher ($453 \pm 35$~cm$^{-1}$) deviates most. Nightingale's spectral slope is the most negative among the four sites, indicating a steeper spectral slope consistent with a finer-grained or more mixed-material surface that dampens Reststrahlen band contrast \cite{Ruff2002BrightData, Shirley2019ParticleEnvironment}.

%Despite these differences, there is some overlap in the spectral parameter ranges among the sites. Distributions of certain band depths (e.g., the $1.4~\mu$m feature) are largely overlapping, indicating those particular spectral characteristics are common across Bennu’s surface. However, other parameters show clear separation. Violin plots of the $2.74~\mu$m hydration band depth (Figure \ref{fig:violin}, left) reveal that Nightingale’s values ($0.1048 \pm 0.0128$) are skewed toward lower band depths than any other site, and violin plots of CF positions (Figure \ref{fig:violin}, right) illustrate that Kingfisher’s CF is consistently at higher frequencies (shorter wavelength) with very little overlap with the others. These observations suggest that while Bennu’s overall composition is relatively uniform, each candidate site exhibits unique spectral nuances that are quantitatively different in key mineralogical indication regions from the others.

\subsection{Principal Component Analysis}

\begin{figure}
    \centering
    \includegraphics[width=1\linewidth]{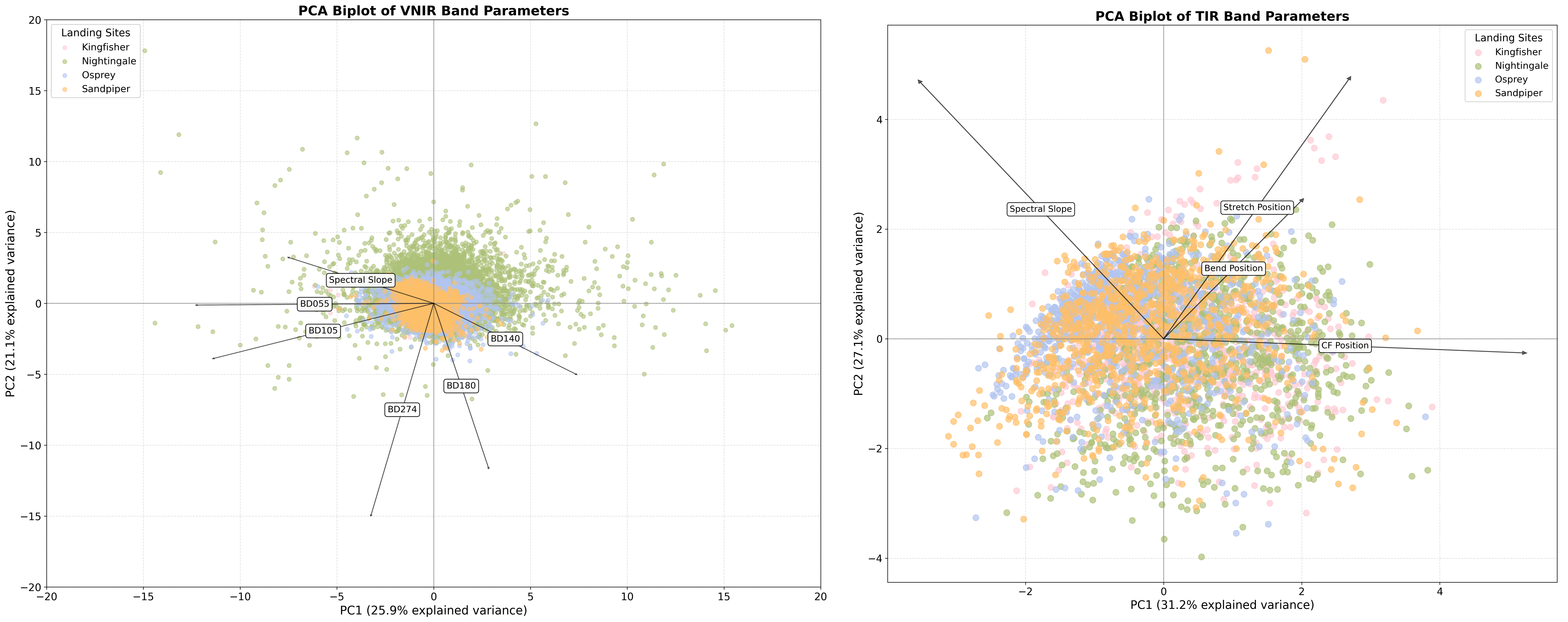}
    \caption{PCA biplots, categorized by candidate site. Vectors describe the influence of each band parameter (i.e., their contributions to PC1 and PC2). (Left): OVIRS data. Most observations are concentrated at the origin, with Osprey and Sandpiper diffused towards the left. Kingfisher is tightly clustered around the origin. Nightingale's more pronounced spread encompasses the variation of the three other sites. (Right): OTES data. Compared to the VNIR results, the TIR observations are more concentrated and all four sites generally exhibit similar spread in all directions, with minor deviations. Sandpiper is shifted slightly to the left, whereas Nightingale is skewed to the right.}
    \label{fig:biplot}
\end{figure}

To further quantify the variance in Bennu’s spectral data and explore multivariate differences between sites, a PCA was conducted on the OVIRS and OTES spectral band parameters separately, condensing the information into a few composite variables. The first two principal components (PC1 and PC2) together capture a significant portion of the dataset's variance (approximately $\sim47\%$ of total variance for each set). PC1 accounts for the largest share (around $\sim26\%$ for both datasets), and is dominated by contributions from broadband spectral slope and overall emissivity or relative reflectance level—effectively distinguishing spectra by their overall brightness and tilt. PC2 (approximately $\sim21\%$ and $\sim27\%$ of variance for OVIRS and OTES, respectively) is influenced by specific reflectance or emission features, particularly the $2.74~\mu$m hydration band and the Restrahlen band depths, respectively. Hence, PC1 can be interpreted as separating spectra by general albedo or temperature-related shape, whereas PC2 separates by compositional features like hydration and silicate mineral signatures.

When plotted in the PC1–PC2 space, the spectra form clusters that correlate with their site of origin (Figure \ref{fig:biplot}). Each site’s data occupies a characteristic region of the scatter plot. For the OVIRS dataset, the PCA (Figure \ref{fig:biplot}, Left) reveals distinct spectral signatures among the four candidate sites. Kingfisher's points cluster closely near the origin, reflecting its overall spectral homogeneity and subdued band depths. Sandpiper's cluster extends along positive PC1, driven by stronger absorptions near $2.74~\mu$m (mean = 0.1127) — indicating enhanced hydration signatures. Osprey's data are dispersed primarily along PC1, influenced by variations in the $1.4~\mu$m and $1.8~\mu$m bands, with intermediate BD274 values (mean = 0.1102) suggesting compositional diversity in its regolith. Nightingale's spectra form a distinct group with the weakest $2.74~\mu$m absorption (mean = 0.1048) but more variable spectral slopes. Formal testing via PERMANOVA \cite{Anderson2001AVariance} and PERMDISP \cite{Anderson2006Distance-basedDispersions} on the full VNIR feature space (999 permutations) confirms that all pairwise site comparisons are significant ($p = 0.001$), but reveals that the dispersion statistic ($F = 1820.2$) substantially exceeds the centroid statistic ($F = 322.0$). This result indicates that inter-site differences in VNIR are driven primarily by within-site heterogeneity rather than systematic shifts in mean band parameters: Kingfisher is tightly clustered while Nightingale's spectra span the full range occupied by all other sites, accounting for the large dispersion contrast (Figure~\ref{fig:biplot}).

Within the OTES dataset (Figure \ref{fig:biplot}, Right), Kingfisher’s spectra lie distinctly toward the high end of PC1, indicating their relatively steep spectral slopes and higher CF position, and cluster tightly, reflecting more homogeneous spectral properties. Osprey’s spectra spread across a broad range on PC2 but at relatively low PC1, suggesting a mix of sub-populations; some Osprey spectra have redshifted CF positions (lowering PC1) whereas others are closer to the global average. Sandpiper’s cluster occupies the lower-PC1, mid-PC2 quadrant, consistent with its more neutral spectral slope and moderate band features. Nightingale’s spectra span a wider area in this space: they have generally lower PC1 scores (flatter slopes) but a wide spread in PC2, reflecting variability in their Si-O bend positions. Notably, Nightingale points extend to the highest PC2 values of any site while also overlapping with the lower PC2 range of Osprey and Sandpiper. This overlap indicates that, whereas Nightingale includes some of the most distinctive spectra, it also shares commonalities with all the other sites. PERMANOVA \cite{Anderson2001AVariance} and PERMDISP \cite{Anderson2006Distance-basedDispersions} applied to the full four-parameter TIR feature space confirm all pairwise comparisons are significant ($p = 0.001$), and here the centroid statistic ($F = 93.5$) is approximately three times larger than the dispersion statistic ($F = 27.4$) --- the inverse of the VNIR pattern. This result indicates that in TIR feature space the sites occupy more distinct compositional positions, consistent with the systematic shifts in CF position and Reststrahlen stretch reported in Table~\ref{tab:band_param_statistics}. Together, the VNIR and TIR results indicate that the two datasets capture complementary aspects of inter-site variation: VNIR distinctiveness is driven primarily by within-site spectral diversity, while TIR separation reflects more systematic differences in bulk composition and particle size. 

\subsection{Cluster Analysis}

\begin{figure}
    \centering
    \includegraphics[width=1\linewidth]{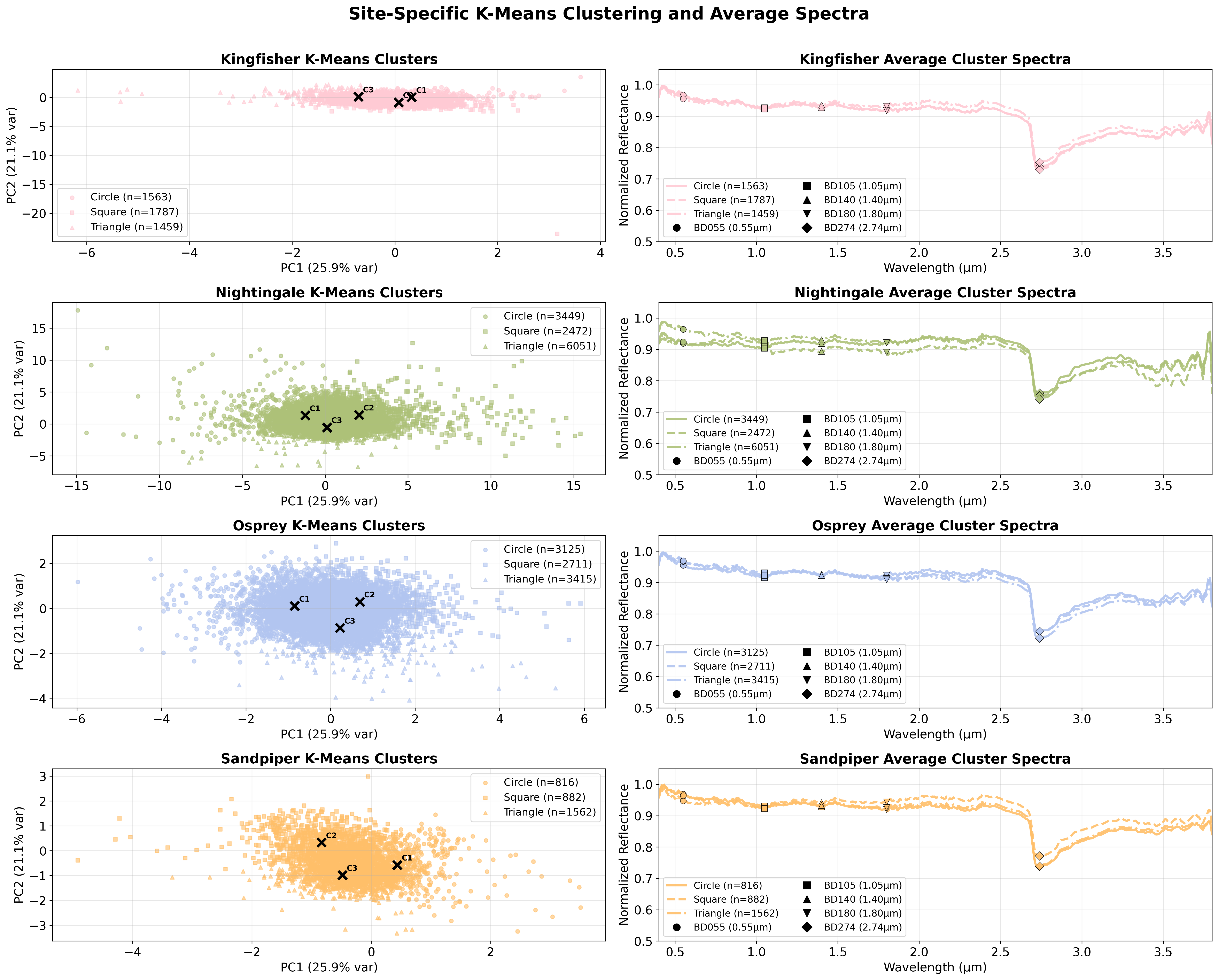}
    \caption{K-means clustering for normalized VNIR (OVIRS) data highlighting within-site variation, demonstrated by the spectral coupling. Nightingale particularly demonstrates distinct sub-grouping with varied $3.5~\mu$m features. Generally, a main population (more than $40\%$ of the site's total spectra) emerges, with two secondary subgroups.}
    \label{fig:vnirknn}
\end{figure}

\begin{figure}
    \centering
    \includegraphics[width=1\linewidth]{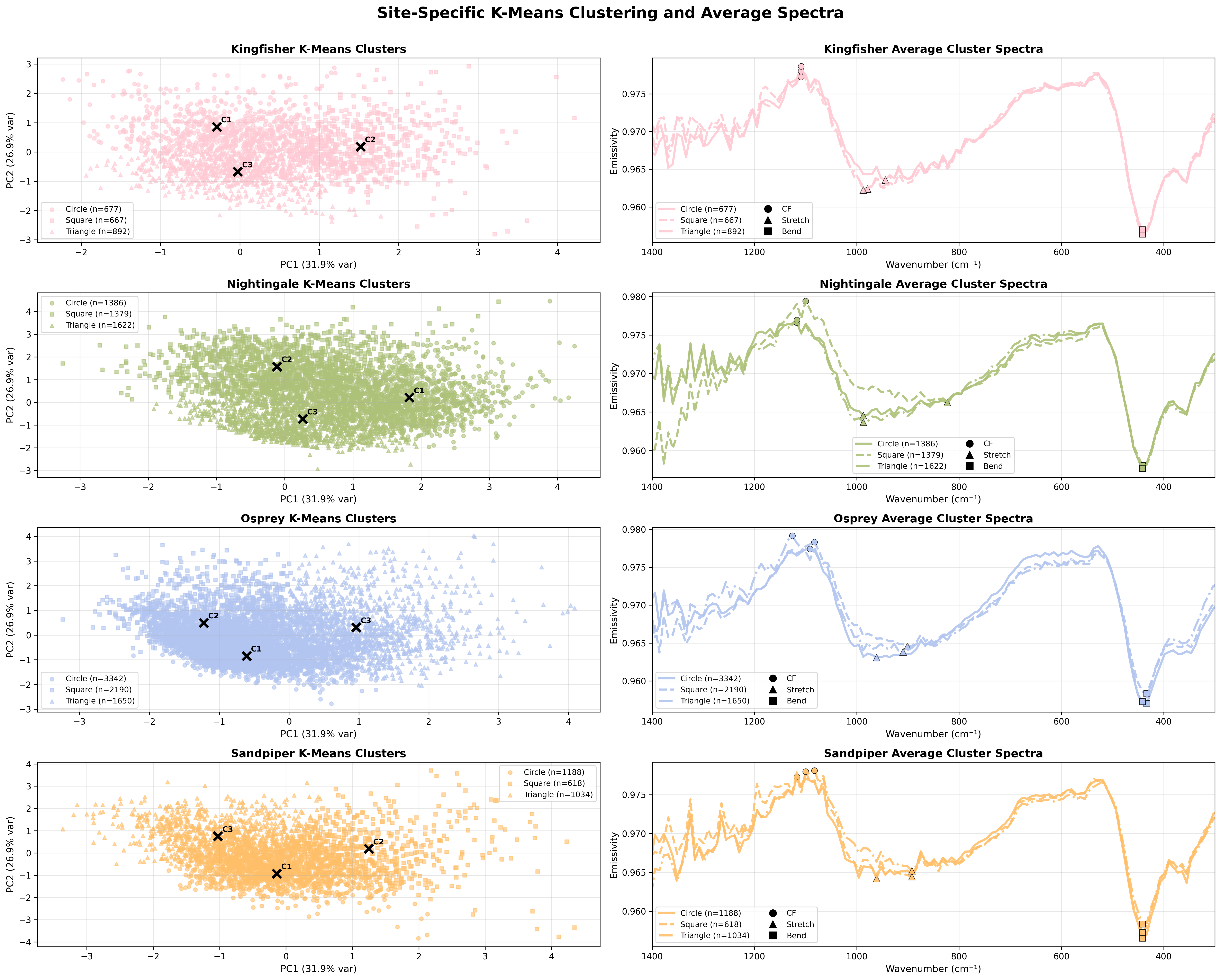}
    \caption{K-means clustering for normalized TIR (OTES) highlighting within-site variation, demonstrated by the spectral coupling. Nightingale contains subgroups of varying spectral slopes and distinct CF feature positions. Populations in TIR are generally more evenly distributed, with the exception of Osprey's main population ($\sim46\%$ of the site's total spectra).}
    \label{fig:TIRknn}
\end{figure}

Using the PCA results as a reduced feature set, we applied K-means clustering to identify natural groupings of spectra, both across and within the candidate sites. In the full-dataset clustering, the algorithm primarily grouped spectra by site, reinforcing that site-to-site differences dominate the variance. However, to investigate within-site heterogeneity, we also performed separate clustering for each site’s spectra. For each sampling site, a k-means solution with $k = 3$ clusters was explored; in all cases, we found the existence of at least two distinct spectral sub-groups per site (Figures \ref{fig:vnirknn} and \ref{fig:TIRknn}).

In the OVIRS analysis (Figure \ref{fig:vnirknn}), Nightingale's spectra were notably more diffuse across the PCA and contained a distinct grouping with distinct behaviour around $3.5~\mu$m. Both Kingfisher and Sandpiper demonstrated similar cluster behaviour, with outliers correlating to slightly brighter surfaces, with higher normalized reflectance spectra. Osprey showed relatively tight clustering, with some diffusion towards the lower PC1 axis indicative of a subpopulation with a slightly deeper hydration feature.

On the other hand, within the OTES dataset (Figure \ref{fig:TIRknn}), Osprey’s spectra (Figure \ref{fig:TIRknn}, Blue) could be divided into a main cluster and one small cluster ($\sim10\%$ of Osprey spectra) that exhibited slightly lower emissivity in the $800$–$1000$ cm$^{-1}$ range (shallower Restrahlen features) and higher PC2 scores. Kingfisher’s spectra (Figure \ref{fig:TIRknn}, Pink) also split into a dominant cluster and a minor cluster distinguished by a steeper long-wavelength slope. Nightingale's subgroups (Figure \ref{fig:TIRknn}, Green) are relatively evenly divided, with the circle and triangle spectra closely resembling one another. The third subpopulation, however, has a steeper spectral slope, with the stretch minima shifted to smaller wavenumbers. Finally, Sandpiper's spectra remain relatively consistent across all three subgroups, with minor variations in the spectral slope and stretch positions. 

The identification of these intra-site clusters underscores that each site, though defined within a small region, contains a diversity of spectral properties. VNIR is more diagnostic of hydrated and ferrous minerals, whereas the TIR additionally considers the particle size (through the slope), space weathering (through the \citeA{Clark2023OverviewBennu} typologies), and relative abundance of phyllosilicates versus anhydrous silicate phases (through stretch and bend positions) \cite{Farmer1975Infra-redChemistry, Ramsey1998MineralSpectra}. Additionally, because VNIR reflectance follows non-linear mixing \cite{Hapke1981BidirectionalTheory}, spectrally active trace minerals can dominate the signal disproportionate to their true abundance. In contrast, TIR emission approximates linear mixing of all mineral components \cite{Ramsey1998MineralSpectra}, more faithfully reflecting bulk mineralogy.

\subsection{Statistical Significance of Spectral Variations}
\label{sec:statsig}

\begin{table}[htbp]
\centering
\caption{Statistical Analysis and Effect Sizes of Spectral Parameters Across Sites}
\label{tab:statistical_analysis}
\resizebox{\textwidth}{!}{%
\begin{tabular}{llccccccc}
\hline
\textbf{Test Type} & \textbf{Comparison} & \textbf{Statistic} & \textbf{df} & \textbf{$\Delta$ BD274} & \textbf{Cohen's $d$} & \textbf{Overlap \%} & \textbf{$\Delta$ CF (cm$^{-1}$)} & \textbf{p-value} \\
\hline
\multicolumn{9}{l}{\textit{\textbf{Welch's One-Way ANOVA: Overall Site Differences}}} \\
\hline
Welch's ANOVA & BD274 ($2.74~\mu$m) & $F = 727.5$ & $(3,\,11314)$ & -- & -- & -- & -- & $< 10^{-16}$ \\
Welch's ANOVA & CF Position & $F = 314.7$ & $(3,\,8281)$ & -- & -- & -- & -- & $< 10^{-16}$ \\
\hline
\multicolumn{9}{l}{\textit{\textbf{Post-hoc Pairwise Comparisons (Welch's $t$-test, Bonferroni-corrected, $\alpha = 0.0083$)}}} \\
\hline
Welch's $t$ & Nightingale vs. Kingfisher & $t = -28.4$ & $14413$ & $-0.0045$ & $-0.395$ & 71\% & $+27.6$ & $1.30 \times 10^{-172}$ \\
Welch's $t$ & Nightingale vs. Osprey & $t = -38.1$ & $19811$ & $-0.0053$ & $-0.495$ & 76\% & $+16.6$ & $8.62 \times 10^{-307}$ \\
Welch's $t$ & Nightingale vs. Sandpiper & $t = -43.2$ & $8343$ & $-0.0078$ & $-0.659$ & 68\% & $+8.2$ & $< 10^{-300}$ \\
Welch's $t$ & Kingfisher vs. Osprey & $t = -5.8$ & $9508$ & $-0.0008$ & $-0.104$ & 89\% & $-11.0$ & $6.49 \times 10^{-9}$ \\
Welch's $t$ & Kingfisher vs. Sandpiper & $t = -18.6$ & $6802$ & $-0.0033$ & $-0.426$ & 72\% & $-19.3$ & $1.98 \times 10^{-75}$ \\
Welch's $t$ & Osprey vs. Sandpiper & $t = -15.8$ & $5390$ & $-0.0025$ & $-0.333$ & 79\% & $-8.4$ & $2.88 \times 10^{-55}$ \\
\hline
\multicolumn{9}{l}{\textit{\textbf{Multivariate Hotelling's $T^2$ Tests (4D Feature Space: BD274, BD140, CF, Stretch)}}} \\
\hline
$T^2$ & Nightingale vs. Kingfisher & $F = 59.7$ & $(4, 1995)$ & \multicolumn{4}{c}{Multivariately distinct} & $< 10^{-16}$ \\
$T^2$ & Nightingale vs. Osprey & $F = 42.4$ & $(4, 1995)$ & \multicolumn{4}{c}{Multivariately distinct} & $< 10^{-16}$ \\
$T^2$ & Nightingale vs. Sandpiper & $F = 94.7$ & $(4, 1995)$ & \multicolumn{4}{c}{Multivariately distinct} & $< 10^{-16}$ \\
$T^2$ & Kingfisher vs. Osprey & $F = 60.9$ & $(4, 1995)$ & \multicolumn{4}{c}{Multivariately distinct} & $< 10^{-16}$ \\
$T^2$ & Kingfisher vs. Sandpiper & $F = 91.3$ & $(4, 1995)$ & \multicolumn{4}{c}{Multivariately distinct} & $< 10^{-16}$ \\
$T^2$ & Osprey vs. Sandpiper & $F = 5.8$ & $(4, 1995)$ & \multicolumn{4}{c}{Multivariately distinct} & $1.32 \times 10^{-4}$ \\
\hline
\end{tabular}%
}
\begin{tablenotes}
\small
\textbf{Note:} Welch's one-way ANOVA confirms that spectral parameters differ significantly among sites: BD$_{2.74}$ ($F_{3,\,11314} = 727.5$, $p < 10^{-16}$) and CF position ($F_{3,\,8281} = 314.7$, $p < 10^{-16}$); degrees of freedom use the Welch--Satterthwaite approximation. Post-hoc pairwise Welch's $t$-tests with Bonferroni correction identify which specific site pairs differ (all statistically significant at $\alpha = 0.0083$). Mean differences shown as $\Delta$ values: negative BD274 indicates deeper band depth for the second site; for CF Position, positive values indicate higher wavenumber for the first site. Cohen's $d$ quantifies standardized effect size: $|d| < 0.2$ = negligible, 0.2--0.5 = small, 0.5--0.8 = medium, $> 0.8$ = large. Overlap \% indicates distribution overlap (higher = less distinct). Multivariate Hotelling's $T^2$ tests evaluate whether sites are distinguishable in combined 4D feature space (BD274 and BD140 from VNIR; CF and Reststrahlen stretch positions from MTIR). Whereas all tests show statistical significance, effect sizes reveal that practical significance varies considerably across site pairs.
\end{tablenotes}
\end{table}

Levene's test confirmed significant heteroscedasticity across all band parameters, hence necessitating the use of Welch's, rather than conventional, ANOVA \cite{Welch1951OnApproach}. Welch's ANOVA confirms that spectral parameters differ significantly among sites: BD274 ($F_{3,11314} = 727.5$, $p < 10^{-16}$) and CF position ($F_{3,8281} = 314.7$, $p < 10^{-16}$). However, with sample sizes of 3,260--11,972 spectra per site (VNIR) and 2,993--8,272 (TIR), such extremely small p-values are expected even for modest differences \cite{Lin2013TooProblem.}.

To assess practical significance, we calculated Cohen's $d$, a standardized effect size measure \cite{Cohen1988StatisticalScience}. For BD274, effect sizes varied considerably (Table~\ref{tab:statistical_analysis}): the largest was Nightingale vs. Sandpiper ($d = -0.659$, medium-to-large), indicating Sandpiper's hydration band is consistently deeper. Small-to-moderate effects were found for Nightingale vs. Osprey ($d = -0.495$), Kingfisher vs. Sandpiper ($d = -0.426$), Nightingale vs. Kingfisher ($d = -0.395$), and Osprey vs. Sandpiper ($d = -0.333$). In contrast, one pair showed negligible effects despite highly significant p-values: Kingfisher vs. Osprey ($d = -0.104$, 89\% distribution overlap) are nearly indistinguishable in BD274.

For CF position, Kingfisher exhibits systematically lower wavenumbers (1093.0 cm$^{-1}$) compared to all other sites (1104--1120 cm$^{-1}$), representing shifts of 11--28 cm$^{-1}$. Post-hoc tests confirm Kingfisher's CF differs significantly from all sites ($p < 10^{-23}$), representing TIR emissivity variations.

Multivariate Hotelling's $T^2$ tests were conducted in a four-dimensional feature space selected to span the primary physical axes of surface variation identified across both instruments: BD$_{2.74}$ and BD$_{1.40}$, representing two spectrally distinct OH/H$_2$O absorption bands that together constrain the hydration state of surface materials; and CF position and Reststrahlen stretch minimum, which jointly constrain silicate bulk composition and particle size in the TIR. Together these four parameters capture complementary information not reducible to any single band parameter. We note that OVIRS and OTES observations were not spatially co-registered at the individual spectrum level (sample sizes differ by site and instrument; Table~\ref{tab:data_overview}), so cross-instrument feature vectors were formed by treating VNIR and TIR parameters as independent site-level distributions rather than matched observations. This approximation is supported by the broadly comparable spatial footprints of both instruments during Reconnaissance ($\sim$2--10~m), but means that fine-scale spatial covariance between VNIR and TIR features within a site cannot be assessed. With this caveat, the $T^2$ results confirm all sites are multivariately distinct ($p \ll 0.01$). Critically, this approach captures distinctions not evident univariately: site pairs similar in BD$_{2.74}$ alone (e.g., Kingfisher vs.\ Osprey) can be distinguished when multiple features are considered jointly. This result is visually evident in PCA biplots (Figure~\ref{fig:biplot}), where sites occupy distinct principal component space regions.

\section{Discussion}

\subsection{Statistical Model Validation}
\label{sec:valid}

Because this analysis depends on combining results from multiple statistical models and interpreting them in the context of very large sample sizes, an assessment of methodological limitations is critical.

\subsubsection{The Large Sample Problem and Effect Size Interpretation}

With sample sizes ranging from thousands to tens of thousands of spectra per site (3,260--11,972 for VNIR; 2,993--8,272 for TIR), our analysis encounters the ``large sample problem'', wherein even trivial differences between group means become statistically detectable at arbitrarily high confidence levels \cite{Lin2013TooProblem., Sullivan2012Enough}. This problem is evident in p-values that approach computational limits (e.g., $p < 10^{-16}$ for CF position ANOVA), which, while confirming that observed differences are not due to chance, provide no information about whether those differences are scientifically meaningful.

Effect size calculations (Cohen's $d$) for all pairwise comparisons reveal that statistical significance does not uniformly translate to practical importance. For example, Osprey vs. Sandpiper yields $p = 2.88 \times 10^{-55}$ for BD274 differences, yet Cohen's $d = -0.333$ (small effect) and 79\% distribution overlap indicate the sites are nearly indistinguishable on this parameter. Conversely, Nightingale vs. Sandpiper shows both high statistical significance and a moderate-to-large effect size ($d = -0.659$), representing a meaningful difference. This discordance between p-values and effect sizes is characteristic of large-sample datasets and underscores the necessity of reporting both metrics \cite{Wasserstein2016ThePurpose}. 

\subsubsection{ANOVA False Results}

Welch's one-way ANOVA relaxes the equal-variance assumption confirmed necessary by Levene's test (Supplementary Information), adjusting degrees of freedom via the Welch--Satterthwaite approximation rather than assuming $\mathrm{MS}_\mathrm{W}$ is homogeneous across groups. However, the fundamental large-sample limitation remains: the F-statistic is still produced by partitioning total variability into between- and within-group 
components,

\[
F = \frac{\mathrm{MS}_\mathrm{B}}{\mathrm{MS}_\mathrm{W}},
\]

where $\mathrm{MS}_\mathrm{B}$ (between-groups mean square) quantifies how much group means deviate from the overall mean, and $\mathrm{MS}_\mathrm{W}$ (within-groups mean square) reflects variability within each group. When sample sizes are highly unbalanced -- as in our case, where Nightingale has nearly four times as many spectra as Sandpiper (11,972 vs. 3,260 for VNIR) -- subtle shifts in the large group's mean disproportionately inflate $\mathrm{MS}_\mathrm{B}$, thereby increasing the F-statistic. To ensure the quality of our conclusions, the F-statistic is supplemented by additional robust analysis including the Cohen's \textit{d} calculations.

\subsubsection{Multivariate Analysis and Joint Interpretation}

Multivariate Hotelling's $T^2$ tests address some limitations of univariate comparisons by evaluating whether sites differ in a combined feature space. Importantly, this approach captures complementary information: site pairs that overlap substantially on one parameter (e.g., Kingfisher vs. Osprey on BD274, 89\% overlap) can still be distinguished using the full suite of features. The F-statistics from Hotelling's tests provide multivariate effect size analogs, with Nightingale vs. Sandpiper ($F = 94.7$) and Kingfisher vs. Sandpiper ($F = 91.3$) showing the strongest separation.

\subsection{Comparison with Global-Scale Statistical Analyses}
In contrast to past global statistical studies, including \citeA{Barucci2020OSIRIS-RExStatistics}, our site-resolved analysis ($\sim$2--10~m per spot) incorporates both OVIRS and OTES observations acquired during the Reconnaissance phase and applies formal inferential testing to quantify spectral heterogeneity at sub-regional scales. Welch's ANOVA ($F_{3,11314} = 727.5$, $p \ll 0.01$) and multivariate Hotelling's~$T^2$ tests ($p \ll 0.01$) demonstrate that the four candidate sampling sites are statistically distinct in both VNIR and TIR band parameter space. PCA and K-means analyses further reveal intra-site sub-populations that reflect compositional and textural variability not resolvable in global datasets.

These results refine and extend the conclusions of \citeA{Barucci2020OSIRIS-RExStatistics}, showing that whereas Bennu is spectrally homogeneous at global scales, statistically significant heterogeneity emerges when examined at the spatial resolution of the candidate sites across a larger spectral range (i.e., OVIRS and OTES). The differences in hydration band depth, CF position, and silicate stretching modes indicate that local surface processes -- such as grain-size sorting, space weathering, and small-scale compositional abundances -- produce measurable spectral  variability at the 2-10 m scale.

\subsection{Site Mineralogical and Physical Differences}
\label{sec:sitemin}
While it is difficult to decouple physical effects (including space weathering and particle size effects) from mineralogical differences as demonstrated by the significant overlap in indicators in Table \ref{tab:bandparam_summary}, some hypotheses can be weighed to draw preliminary conclusions. Band parameters used in this work represent a deliberate simplification of the full spectral information to assess our statistical classification pipeline. The interpretations offered here are informed by, and consistent with, qualitative examination of the site-mean spectra (Figure~\ref{fig:site_spectra}), but a rigorous test of these hypotheses would require full spectral shape analysis or spectral unmixing applied to the complete OVIRS and OTES datasets --- an avenue we identify as a priority for future work.

\subsubsection{Surface Roughness and the VNIR Spectral Slope}
The ANOVA of the OVIRS data resulted in the spectral slope having the third largest f-statistics, with a value of 509.8 and statistically significant. Investigations into the phase reddening on Bennu found a monotonically decreasing spectral slope, consistent with other B-type asteroids \cite{Fornasier2020PhaseSpectroscopy}. Despite this consistent global behaviour, each of the four sites has a characteristic slope. Some of the four sites have steeper reddening coefficients (or slopes) than the global average, which could indicate the presence of a thin, dust layer with micro-roughness influencing these measurements \cite{Fornasier2020PhaseSpectroscopy}. This fine dust is also consistent with TIR spectral slope variations observed across the sites: \citeA{Hamilton2021EvidenceSpectroscopy} demonstrated that the OTES spectral slope across Bennu is diagnostic primarily of fine particle abundance, and the relatively lower thermal inertia at Nightingale \cite{Rozitis2022High-ResolutionBennu} provides independent thermophysical support for fine-grained material at that site.

\subsubsection{Magnetite Features}
In the VNIR data, both the BD055 and BD180 parameters indicate the presence of magnetite \cite{Simon2020WeakSpectrometer,Lauretta2024AsteroidOSIRIS-REx}. Remote sensing investigations have indicated a correlation between surface variation features, including surface albedo, and the depth of the 0.55 $\mu$m band \cite{Simon2020WeakSpectrometer}, which is usually associated with the iron transition band \cite{Simon2020WeakSpectrometer}. Nightingale ($-0.0038 \pm 3.5\times10^{-7}$), Kingfisher ($-0.0032 \pm 2.5\times10^{-7}$), and Osprey ($-0.0035 \pm 3.0\times10^{-7}$) have similar band depths, while Sandpiper ($-0.0027 \pm 2.8\times10^{-7}$) is less pronounced. The presence of magnetite is an indicator of extensive aqueous alteration \cite{Hamilton2019EvidenceBennu}, which is consistent with the finding of bright veins across the surface of Bennu \cite{Kaplan2020BrightHistory} and the presence of phyllosilicates \cite{Hamilton2019EvidenceBennu}. The difference in these band depths (Figure \ref{fig:violin}) could indicate variability in the extent of alteration at Sandpiper.

\subsubsection{Particle Size Effects and the TIR Spectral Slope}
Past investigations \cite{Ruff2002BrightData, Shirley2019ParticleEnvironment, DonaldsonHanna2019SpectralStudy} have determined that TIR spectral slope acts as a good proxy to determine trends in particle size. As a general rule, the steeper the slope, the smaller the particle size. Other effects of particle size include a change in spectral contrast, depending on the environmental conditions of the measurement and thermal gradients present. In this case, in Bennu's environment, we note that Kingfisher has more subdued RBs and Sandpiper more pronounced, which could indicate a compartively larger amount of fine material at Sandpiper \cite{Shirley2019ParticleEnvironment}. 

\citeA{Burke2021ParticleBennu} determined approximate particle size frequency distributions (PSFD) of each of the four sites which, in conjunction with this work, allows us to make some conjectures about the sites. We note, however, that this comparison spans different spatial scales: OTES integrates over a footprint of approximately 2--10~m \cite{Christensen2018TheInstrument}, providing an area-averaged measure of the surface particle environment, whereas the \citeA{Burke2021ParticleBennu} PSFDs are derived from OCAMS images acquired at 0.004--0.01~m~pixel$^{-1}$ during Reconnaissance, resolving individual centimeter-scale particles. Focusing on the spectral slope, we see from Figure \ref{fig:violin} and Table \ref{tab:band_param_statistics} that the slopes are, in increasing order, Sandpiper, Osprey, Nightingale, and Kingfisher. In other words, we would expect for the particle size to increase along that order as well. It is important to note that Sandpiper and Osprey are fairly similar, whereas Kingfisher and Nightingale are distinct from the first pair and close together. This result is supported by the findings in \citeA{Burke2021ParticleBennu}, which states that the minimum size estimate for Sandpiper and Osprey are both 0.01 m, whereas Kingfisher and Nightingale are both 0.02 m. Although the particle sizes expected to impact spectral slope are on the micron scale, the trend appears to be the same, potentially indicating fractal fragmentation of boulders \cite{Hsu2022Fine-grainedAsteroids}. This trend can be further explored through experimentation into the surface versus volume scattering effects caused by dust layering on porous surfaces, hypothesized by \citeA{Biele2019EffectsBodies, Rozitis2022High-ResolutionBennu}.

\subsubsection{OTES Spectral Types}

\citeA{Clark2023OverviewBennu} define three OTES spectral classifications for Bennu based on the position of the silicate stretching band minimum, building on earlier work by \citeA{Lantz2020CanData} and \citeA{Hamilton2021EvidenceSpectroscopy}:

\begin{enumerate}
    \item \textbf{Type~1}: emissivity minimum near $987$~cm$^{-1}$.
    \item \textbf{Type~2}: emissivity minimum near $814$~cm$^{-1}$.
    \item \textbf{Type~3}: emissivity minimum near $920$~cm$^{-1}$
\end{enumerate}

\noindent The origin of these spectral types remains an open question. 
\citeA{Clark2023OverviewBennu} note that Type~3, lying between the two 
end-member positions, may reflect subpixel mixing of Type~1 and Type~2 
material rather than a distinct surface process. Until the distribution of these types is examined as a function of thermal properties, viewing geometry, and geographic context, a space weathering interpretation cannot be confirmed, and we therefore treat this classification as descriptive rather than causal.

As shown in Table~\ref{tab:band_param_statistics}, the mean stretching band positions place 
Kingfisher ($988 \pm 72$~cm$^{-1}$) and Nightingale ($963 \pm 65$~cm$^{-1}$) nearer to the Type~1 reference ($\sim$987~cm$^{-1}$), and Osprey ($947 \pm 65$~cm$^{-1}$) and Sandpiper ($941 \pm 67$~cm$^{-1}$) nearer to Type~3 ($\sim$920~cm$^{-1}$). However, the within-site standard deviations ($\sim$65--72~cm$^{-1}$) are large relative to the separation between type centres ($\sim$647~cm$^{-1}$), meaning the site distributions overlap substantially and individual spectra at each site span both type regions. This classification therefore reflects the dominant spectral character of each site's mean: the Welch's $t$-test confirms that site mean positions are statistically distinguishable from one another (all pairwise $p < 0.001$; Table~\ref{tab:statistical_analysis}), but the broad within-site variance --- reflected in the K-means sub-populations (Figure \ref{fig:TIRknn}) --- indicates that both spectral types likely coexist within each site at the scale of individual OTES footprints. Interestingly, this two-pair grouping mirrors the site pairings identified in the particle size analysis of \citeA{Burke2021ParticleBennu}, suggesting that the stretch position differences may reflect coupled surface properties rather than a single driving process.

\subsubsection{CF and Other Silicate Indicators}
The shifts in CF position to shorter wavelengths could be an indication of decreased Fe and increased Mg content in the silicates present in the sites \cite{Bates2020LinkingReturn}. It can also be an indication of increased aqueous alteration, as past investigations into CM and CI meteorites has shown that an increase in Mg content is indicative of increased aqueous alteration \cite{Tomeoka1985IndicatorsNi,Bates2020LinkingReturn}.  

\citeA{Hamilton2021EvidenceSpectroscopy} found the position of the silicate bend minima was found to be at 440 cm$^{-1}$ for the global spectra. Hence, Osprey and Sandpiper, with 441 $\pm$ 24 cm$^{-1}$ and 438 $\pm$ 11 cm$^{-1}$ respectively, are the two closest sites to the global spectra, followed by Nightingale (447 $\pm$ 29 cm$^{-1}$) and Kingfisher (453 $\pm$ 35 cm$^{-1}$). Once again, we note that the sites are coupled in the same way as the two previous groupings. Interestingly, the standard deviation for the bend locations was notably smaller for all four sites compared to the other band parameters. 

The global CF feature, as identified in \citeA{Hamilton2021EvidenceSpectroscopy}, was found to be at 1090 cm$^{-1}$. That would place Kingfisher as the closest (1093 $\pm$ 56 cm$^{-1}$), followed by Osprey (1104 $\pm$ 41 cm$^{-1}$), Sandpiper (1112 $\pm$ 43 cm$^{-1}$), and Nightingale (1121 $\pm$ 43 cm$^{-1}$). Unlike for the silicate bands, Kingfisher's position is closest to that of the global spectra. However, all four sites remain within one standard deviation of the global CF position of 1090~cm$^{-1}$ \cite{Hamilton2021EvidenceSpectroscopy}, and the site ranges overlap substantially. The observed ordering is based on site mean positions, which are statistically distinguishable (Welch's $t$-test, $p < 0.001$), but should not be interpreted as indicating that sites occupy categorically distinct compositional regimes.

\subsubsection{Evidence of Aqueous and Thermal Alteration}

Although we found the sites to be similar in the VNIR PCA biplot (Figure \ref{fig:biplot}, Left), we can note that there is still a clear difference across the four sites in terms of the depth of the feature at 2.74 $\mu$m (Figures \ref{fig:site_spectra} and \ref{fig:violin}), confirmed by its band parameter having the largest F-statistic (727.5). The depth of the feature varies across the four sites: in increasing order, Nightingale ($0.1052 \pm 6.9\times10^{-6}$), Kingfisher ($0.1095 \pm 1.0\times10^{-5}$), Osprey ($0.1104 \pm 6.9\times10^{-6}$), and Sandpiper ($0.1127 \pm 1.1\times10^{-5}$). The feature, in analogue materials, is due to structural hydroxyl ions in hydrous clay minerals, common in CI and CM meteorites \cite{Farmer1975Infra-redChemistry, Cloutis2011SpectralChondrites,Hamilton2019EvidenceBennu, Bates2020LinkingReturn, DonaldsonHanna2021SpectralBodies}. The position of the minimum can also shift with mineral structure and composition \cite{Hamilton2019EvidenceBennu}.

\subsubsection{Assessing Nightingale's Representativeness}

A retroactive assessment of Nightingale's representativeness of Bennu's small-scale heterogeneity can support future sample selection and return efforts to other destinations. During the selection process, scientific interests were weighed alongside the feasibility of conducting a TAGSAM operation at the location \cite{Lauretta2017OSIRIS-REx:Bennu}. With a retroactive look at the collected data from the Reconnaissance operations, we can see that Nightingale shares characteristics with the other three sites, as depicted in the band parameter distributions shown in Figure \ref{fig:violin}, and the PCA biplots in Figure \ref{fig:biplot}. While Welch's ANOVA and multivariate tests confirm Nightingale is statistically distinct from the other sites (Table \ref{tab:statistical_analysis}), the effect sizes reveal nuanced relationships. Nightingale shows the smallest differences from Kingfisher (Cohen's $d = 0.40$), moderate differences from Osprey ($d = 0.50$), and the largest differences from Sandpiper ($d = 0.66$). Additionally, Nightingale's spectral parameters span an intermediate range: its mean BD274 (0.1048) is the shallowest among the four sites, yet individual Nightingale spectra exhibit the full range of variability observed across the investigated sites. The significantly greater within-site spectral dispersion at Nightingale ($F = 1820.2$, $p = 0.001$) relative to the other sites independently corroborates this interpretation: Nightingale's surface encompasses a broader range of compositional and physical states, consistent with its location near Bennu's north polewhere a diversity of boulder types and fine material accumulations have been documented \citeA{Dellagiustina2020VariationsBennu}. This finding suggests that whereas Nightingale possesses unique mean characteristics, it also encompasses considerable internal heterogeneity, making it broadly representative of the diversity encompassed by the candidate sites. 

In other words, if another one of the four sites were selected instead of Nightingale as the sample site, the samples would not be as representative of the global behaviour on Bennu and small scale heterogeneity captured in this work, resulting in a skewed understanding of Bennu. The insight in the selection of Nightingale provides us with a relatively ``global'' sample within the context of other sites investigated at this spatial range, allowing us to experiment on a representative understanding of Bennu.

\subsection{Consolidating VNIR and TIR Results}

Using data from both OVIRS and OTES provides complementary sensitivity to different aspects of surface variability. VNIR is dominated by spectrally active phases — particularly the phyllosilicates and hydrated minerals driving the 2.74 µm feature — while TIR approximates a linear mixture of all minerals present, giving broader insight into bulk silicate mineralogy and physical surface properties \cite{Hunt1971AlteredInfrared,Farmer1975Infra-redChemistry, Salisbury1989ThermalSurfaces, Ramsey1998MineralSpectra}. In this dataset, the two instruments are broadly consistent in their site orderings: sites showing stronger hydration signatures in VNIR (Sandpiper and Osprey) also tend to show CF positions and Reststrahlen band structures in TIR consistent with more altered, phyllosilicate-rich assemblages \cite{Hamilton2021EvidenceSpectroscopy}. However, many individual parameters remain partially degenerate: CF position responds to both Mg/Fe ratio and grain size, the TIR spectral slope reflects both particle size and space weathering state, and the 2.74 µm band depth is sensitive to both phyllosilicate abundance and thermal history. The combined dataset therefore narrows, but does not uniquely resolve, the driving factors of inter-site variability. Table \ref{tab:vnir_TIR_site_variability} summarizes which phenomena are constrained by each instrument, reflecting where the two datasets reinforce one another and where ambiguity remain.

\begin{sidewaystable}
\centering
\caption{Drivers of surface variability across Bennu’s four candidate sampling sites as inferred from combined VNIR (OVIRS) and TIR (OTES) observations. VNIR parameters non-linearly weight spectrally active phases, while TIR parameters approximate linear bulk mineral abundances and additionally constrain particle size and space weathering state. Qualitative descriptors summarize relative comparisons between sites based on site-mean band parameters and statistically significant trends defined throughout Sections \ref{sec:statsig} and \ref{sec:sitemin}.}
\label{tab:vnir_TIR_site_variability}
\begin{tabular}{p{4.4cm} p{3.4cm} p{3.6cm} c c c c}
\hline
\textbf{Phenomenon} &
\textbf{VNIR Sensitivity (OVIRS)} &
\textbf{TIR Sensitivity (OTES)} &
\textbf{Nightingale} &
\textbf{Osprey} &
\textbf{Sandpiper} &
\textbf{Kingfisher} \\
\hline

Hydrated phyllosilicates / aqueous alteration
& High sensitivity (2.74~$\mu$m OH band depth; 1.4--1.8~$\mu$m features)
& Moderate sensitivity (CF position; silicate band structure)
& Less altered
& Moderately altered
& Highly altered
& Moderately altered \\

Bulk silicate mineral assemblage
& Low sensitivity; trace and surface-weighted
& High sensitivity (CF position; Reststrahlen bands)
& Similar
& Similar
& Similar
& Distinct \\

Relative mineral abundance (bulk)
& Low; non-linear and dominance-weighted
& High; approximates linear mixing regime
& Similar
& Similar
& Similar
& Similar \\

Particle size / surface texture effects
& Moderate sensitivity (VNIR spectral slope, band contrast)
& High sensitivity (pre-CF slope; emissivity contrast)
& Moderate effects
& Moderate effects
& Less effects
& More effects \\

OTES Spectral Classification
& Indirect (slope reddening; weak feature modulation)
& High sensitivity (Si--O stretch position; \citeA{Clark2023OverviewBennu} typology)
& Type 1
& Type 3
& Type 3
& Type 1 \\
\hline
\end{tabular}
\end{sidewaystable}

\section{Conclusion}

This work provides a quantitative, multi-instrument assessment of spectral variability across the four OSIRIS-REx Reconnaissance candidate sites on Bennu, integrating diagnostic band parameters from OVIRS and OTES with multivariate and inferential statistical analyses. Across both datasets, the sites display subtle but measurable differences in key band parameters. In the VNIR, band depths and spectral slopes vary systematically, with the $2.74~\mu$m OH feature's band depth differing by $\sim$0.01 among sites and with Sandpiper and Osprey exhibiting the strongest hydration signatures. In the TIR, systematic shifts of $\sim$20--30~cm$^{-1}$ in the CF position, silicate stretch and bend positions distinguish the sites, reflecting variations in silicate composition, particle size, and hydration state.

Our multivariate statistical approach combines traditional significance testing (ANOVA, t-tests, Hotelling's $T^2$) with effect size quantification (Cohen's $d$, overlap coefficients) and classification-based validation (PCA, K-means), mitigating the interpretive pitfalls of p-values in large-sample contexts. While univariate comparisons show that practical importance varies considerably -- some site pairs exhibit medium-to-large differences (e.g., Nightingale vs. Sandpiper, $d = -0.659$) whereas others are nearly indistinguishable on single parameters (e.g., Kingfisher vs. Osprey, $d = -0.104$, 89\% overlap in BD274) -- multivariate analysis reveals that all sites are clearly distinct when multiple spectral features are considered together. PCA separates the four sites in low-dimensional feature space, with hydration- and slope-related parameters dominating variance in VNIR and silicate-dependent features governing TIR variability. Hotelling's $T^2$ tests confirm that all pairwise site comparisons are multivariately distinguishable ($p \ll 0.01$), demonstrating that the combined spectral signature of each site is unique. Future work could apply Bayesian hierarchical models that naturally incorporate uncertainty and effect size estimation without reliance on p-values \cite{Kruschke2013BayesianTest.}.

Placed in context of previous global-scale analyses \cite{Barucci2020OSIRIS-RExStatistics, Kaplan2020VisiblenearOSIRIS-REx, Dellagiustina2020VariationsBennu}, which found Bennu to be largely homogeneous at $\sim$20--30~m resolution, these results demonstrate that statistically distinguishable heterogeneity emerges at Reconnaissance-scale ($\sim$2--10~m) sampling.

Some limitations remain. Differences in SNR, viewing geometry, and unbalanced sample sizes introduce uncertainty into the statistical tests. Additionally, several band parameters are influenced by multiple physical processes that remain partially degenerate (e.g., particle size versus surface weathering state). Mineralogical interpretations are also constrained by the use of analogue end-members that may not fully reflect Bennu’s compositional diversity, particularly if Bennu represents alteration states intermediate to or outside the range sampled by the meteorite catalogue. Nevertheless, the convergence of multiple independent statistical approaches builds confidence that the detected variability represents real surface heterogeneity.

Overall, the results demonstrate that Bennu’s surface preserves measurable spectral diversity driven by coupled variations in aqueous alteration, particle size, and mineralogical composition. This heterogeneity indicates that primary parent-body alteration processes were spatially non-uniform, and that secondary surface modification --- including regolith gardening and space weathering --- operates differentially across a body that is globally homogeneous at larger scales. The site-to-site comparisons further highlight the representativeness of Nightingale as the selected sample site: its spectral properties encompass the full range observed across all four locations, capturing both hydrated and less altered endmembers. The framework developed here establishes a quantitative baseline for contextualizing VNIR/TIR observations for laboratory analyses of the returned samples, enabling a more complete interpretation of Bennu’s geological evolution and the processes that shape carbonaceous asteroids.

\section*{Open Research Section}
OSIRIS-REx Thermal Emission Spectrometer and OSIRIS-REx Visible and near-InfraRed Spectrometer datasets from the Detailed Survey and Reconnaissance phases of the Origins, Spectral Interpretation, Resource Identification, and Security–Regolith Explorer mission are available via the Planetary Data System (PDS) \cite{Christensen2018TheInstrument} and \cite{Reuter2018TheBennu}, respectively. OCAMS images from the Reconnaissance phase are also available via the PDS \cite{Rizk2018OCAMS:Suite}. 

\acknowledgments
E. C. Belhadfa acknowledges funding support from the UK Science and Technology Facilities Council (STFC), under studentship grant number 2929131. Her PhD is additionally funded through the University of Oxford's Clarendon Scholarship and Oriel College's Graduate Teaching and Research Fellowship. This work was made possible by the OSIRIS-REx mission and science team, supported by NASA under Contract NNM10AA11C issued through the New Frontiers Program. 

%%%%%%%%%%%%%%%%%%%%%%%%%%%%%%%%%%%%%%%%%%%%%%%
% REFERENCES and BIBLIOGRAPHY
%
% \bibliography{<name of your .bib file>} don't specify the file extension
% don't specify bibliographystyle
%
%%%%%%%%%%%%%%%%%%%%%%%%%%%%%%%%%%%%%%%%%%%%%%%

\bibliography{references.bib}

%Reference citation instructions and examples:
%
% Please use ONLY \cite and \citeA for reference citations.
% \cite for parenthetical references
% ...as shown in recent studies (Simpson et al., 2019)
% \citeA for in-text citations
% ...Simpson et al. (2019) have shown...
%
%
%...as shown by \citeA{jskilby}.
%...as shown by \citeA{lewin76}, \citeA{carson86}, \citeA{bartoldy02}, and \citeA{rinaldi03}.
%...has been shown \cite{jskilbye}.
%...has been shown \cite{lewin76,carson86,bartoldy02,rinaldi03}.
%... \cite <i.e.>[]{lewin76,carson86,bartoldy02,rinaldi03}.
%...has been shown by \cite <e.g.,>[and others]{lewin76}.
%
% apacite uses < > for prenotes and [ ] for postnotes
% DO NOT use other cite commands (e.g., \citet, \citep, \citeyear, \nocite, \citealp, etc.).
%

\end{document}

%% ------------------------------------------------------------------------ %%
%
%  TITLE
%
%% ------------------------------------------------------------------------ %%

%\includegraphics{agu_pubart-white_reduced.eps}

\title{Supporting Information for "Quantifying Surface Heterogeneity Across Asteroid (101955) Bennu using Candidate Site Remote Sensing Data"}
%
% e.g., \title{Supporting Information for "Terrestrial ring current:
% Origin, formation, and decay $\alpha\beta\Gamma\Delta$"}
%
%DOI: 10.1002/%insert paper number here%

%% ------------------------------------------------------------------------ %%
%
%  AUTHORS AND AFFILIATIONS
%
%% ------------------------------------------------------------------------ %%

% List authors by first name or initial followed by last name and
% separated by commas. Use \affil{} to number affiliations, and
% \thanks{} for author notes.
% Additional author notes should be indicated with \thanks{} (for
% example, for current addresses).

% Example: \authors{A. B. Author\affil{1}\thanks{Current address, Antartica}, B. C. Author\affil{2,3}, and D. E.
% Author\affil{3,4}\thanks{Also funded by Monsanto.}}

\authors{E. C. Belhadfa\affil{1}, N. E. Bowles\affil{1}, K. A. Shirley \affil{1}, A. A. Simon \affil{2}, V. E. Hamilton \affil{3}, H. H. Kaplan \affil{2}}

% \affiliation{1}{First Affiliation}
% \affiliation{2}{Second Affiliation}
% \affiliation{3}{Third Affiliation}
% \affiliation{4}{Fourth Affiliation}

\affiliation{1}{Atmospheric, Oceanic and Planetary Physics, Clarendon Laboratory, University of Oxford, Oxford, United Kingdom}
\affiliation{2}{NASA Goddard Space Flight Centre, United States of America}
\affiliation{3}{Southwest Research Institute,  United States of America}
%(repeat as many times as is necessary)

%% ------------------------------------------------------------------------ %%
%
%  BEGIN ARTICLE
%
%% ------------------------------------------------------------------------ %%

% The body of the article must start with a \begin{article} command
%
% \end{article} must follow the references section, before the figures
%  and tables.

\begin{article}

%% ------------------------------------------------------------------------ %%
%
%  TEXT
%
%% ------------------------------------------------------------------------ %

\section{Data Processing Pipeline}

All data used for this work is publicly available on the Planetary Database System (PDS), through the OSIRIS-REx mission bundle. While this work focuses on L3 products for OVIRS and OTES (i.e., processed and calibrated by the instrument science teams), additional insight into the data products through a careful cross-referencing of L2, L1, and L0 products was also required. This supplementary materials outlines the data processing pipeline, including anomalies and artifacts found within the PDS data, required to complete this study.

\subsection{OTES Data}
The OTES data used for this study is the TIR emissivity data (L3 products), available on PDS within the Spectral Analysis V1 package.

\subsubsection{Processing Pipeline}
To produce the dataset used in this study, the L3 PDS products were quality controlled to ensure any findings were the result of compositional or physical drivers, rather than instrument artifacts. Figure \ref{fig:otesdp} outlines the process used for the OTES data.

\begin{figure}
    \centering
    \includegraphics[width=1\linewidth]{OTES Data Pipeline.png}
    \caption{OTES Data Pipeline}
    \label{fig:otesdp}
\end{figure}

\subsubsection{Data Artifacts and Quality Control}
Some off-nominal observations were found within the TIR emissivity dataset. This section briefly outlines the main flags and artifacts encountered and their causes.

\paragraph{Thermal Artifacts}
As expected based on the OTES instrument design \cite{Christensen2018TheInstrument} and normal operation of an FTIR with an uncooled infrared detector, when the scene temperature is close to that of the instrument, there is not enough difference between the two. This similarity limits the instrument's ability to collect a good spectral reading and causes a phase inversion.

\paragraph{Flags and Additional Processing}
Outlined in both the OTES Caveats document available on PDS and in Appendix A.1 of \cite{Hamilton2021EvidenceSpectroscopy}, the OTES data contain quality flags that indicate what data contains thermal artifacts, discussed above, as well as other additional constraints and considerations. Specifically, the quality notes are comprised of a 16-bit flag, indicating the quality of the radiometric measurements and the derived Brightness Temperature (BT). This flag was used to remove known outliers.

\subsection{OVIRS Data}
The OVIRS data used for this study is the VNIR I/F spectra (L3c products), available on PDS within the Spectral Analysis V1 package. This dataset includes selected data, used previously for the Bennu product maps (e.g., \cite{Kaplan2020VisiblenearOSIRIS-REx}). Unlike the OTES dataset, the OVIRS package does not include flags and thus additional consideration and care is required when using large sets of its processed data.

\subsubsection{Processing Pipeline}
To produce the dataset used in this study, the L3 PDS products were quality controlled to ensure any findings were the result of compositional or physical drivers, rather than instrument artifacts. Figure \ref{fig:ovirsdp} outlines the process used for the OVIRS data.

\begin{figure}
    \centering
    \includegraphics[width=1\linewidth]{OVIRS data pipeline.png}
    \caption{OVIRS Data Pipeline}
    \label{fig:ovirsdp}
\end{figure}

\subsubsection{Common Data Artifacts}
\begin{figure}
    \centering
    \includegraphics[width=1\linewidth]{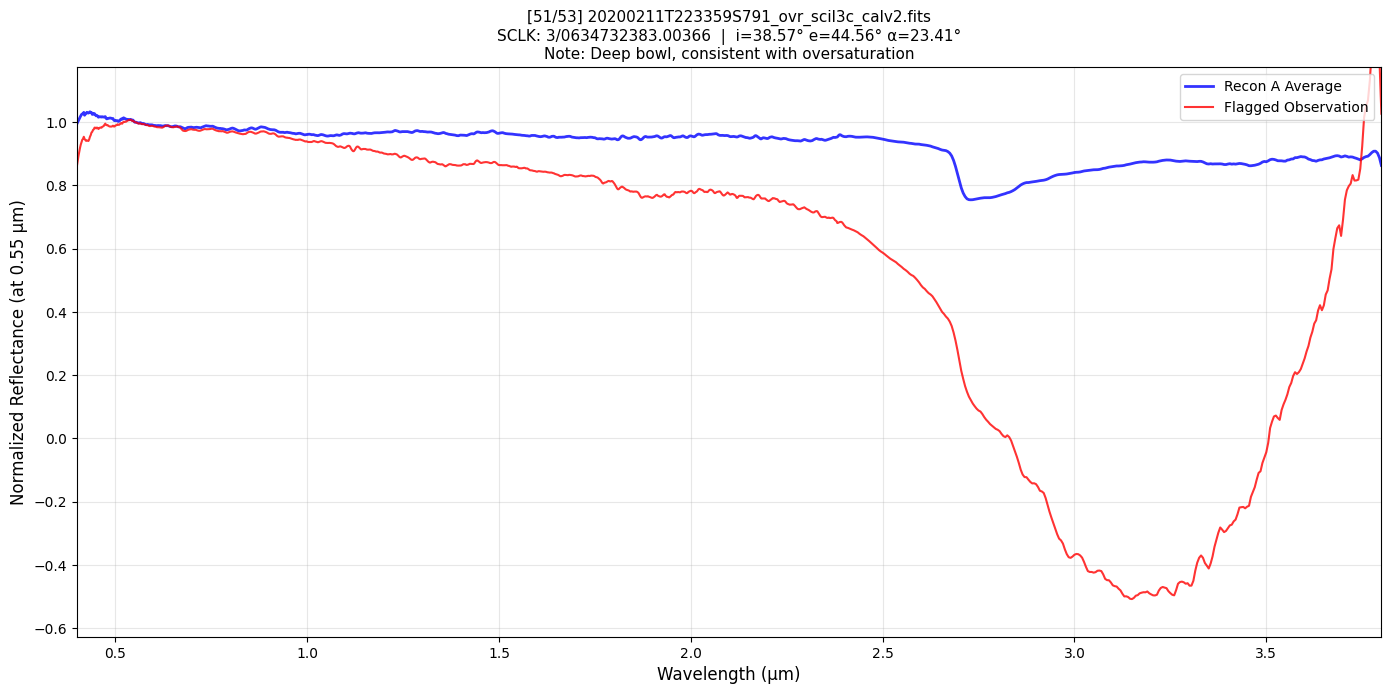}
    \caption{Deep Bowl, Associated with Oversaturation}
    \label{fig:deepoversat}
\end{figure}

When assessing the site-specific datasets (outlined in Table \ref{tab:data_overview}), anomalies were found in Osprey's Reconnaissance B data, as suggested in \cite{Simon2021DerivationCalibration}. Specifically, a large bowl-like shape was found around $2.74 \mu$m (Figure \ref{fig:deepoversat}. Originally thought to be a compositional feature, potentially indicative of increased aqueous alteration, further conversations with the OVIRS instrument team revealed the feature to be an artifact resulting from off-nominal instrument use and, at times, the existing calibration and processing pipeline. To evaluate the extent and drivers of these artifacts, the 11,314 spectra from Osprey's L3c PDS data were manually sorted.

This sorting was accomplished by comparing each observation to the spectra average of Osprey's Recon A observations. They were then manually sorted into one of three categories: 1) nominal, meaning the observation appeared to contain identifiable, expected signatures; 2) artifacts, meaning the observation contained off-nominal signatures and features -- the main of which are outlined here; 3) noisy, meaning the observation was unrecognizable, contained clear random noise, and/or had identifiable repeated noise features. 4,321 observations were found to be nominal, representing $\sim 38\%$ of the L3 data available within the spectral analysis package on PDS. A full list of SCLK identifiers used in this study is available from the first author upon reasonable request. 

The data thought to contain artifacts were further examined through the corresponding resampled spectra (L3a products), thermal tail corrected spectra (L3b) and science observations (L2 products, available within the OVIRS PDS package).

\paragraph{Oversaturation}
During long observation days, such as Osprey's Recon B (2020-02-11), the background level and random noise increase.  At higher background levels pixels could become non-linear in the HgCdTe focal plane, or reach full well value and no longer give the correct counts. Background observations before and after the observation were used to remove the instrument thermal signal and limit the impact of non-linearity.  However, as the instrument warms, the detector sensitivity changes. There are limited accurate radiometric calibration at higher temperatures, causing a disconnect in the calibration process. This oversaturation effect produces a curved, bowl-like continuum at longer wavelengths where the correction amplifies small additive errors.

\paragraph{Out of Band Signal}
Some observations included a repeating pattern caused by an out of band signal from a hot surface, also known as light leaks. The OVIRS pipeline attempts to remove out of band signal by modeling the IR photons and performing a linear fit to ground testing of out of band response.  However, as the IR sensitivity could be reduced for off-nominal temperatures, the signal can be underrepresented and not fully removed. 

\paragraph{Rolling Readout Issues}
When the spacecraft crosses a shadow or is positioned such that the viewing geometry is limb rather than nadir, rolling readout issues can occur resulting in random high or low sections in the spectrum.

\paragraph{Thermal Tail Corrections}
Resampled spectra (L3a) include thermal emission from the surface of Bennu, which must be modelled and removed to extract compositional information. Hence, an automated thermal tail correction, based on measured blackbody fits (L2 products), were applied to create the L3b dataset. However, when the noise or poor calibration were present at longer wavelengths, the automated fit could incorrectly model the thermal information, resulting in secondary peaks, increasing reflection slopes, and/or bowl-like features at longer wavelengths (i.e. $>3\mu$m). 

\paragraph{Instrument Noise}
Finally, like all instruments, OVIRS captured some observations that were compromised by random and instrument noise. Owing to the unpredictability of such noise, these observations must be manually removed. The z-score analysis also aids in removing some more extreme examples that may fall outside three standard deviations of the mean. 

\section{ANOVA Assumptions}
\label{anova}

\begin{figure}
    \centering
    \includegraphics[width=1\linewidth]{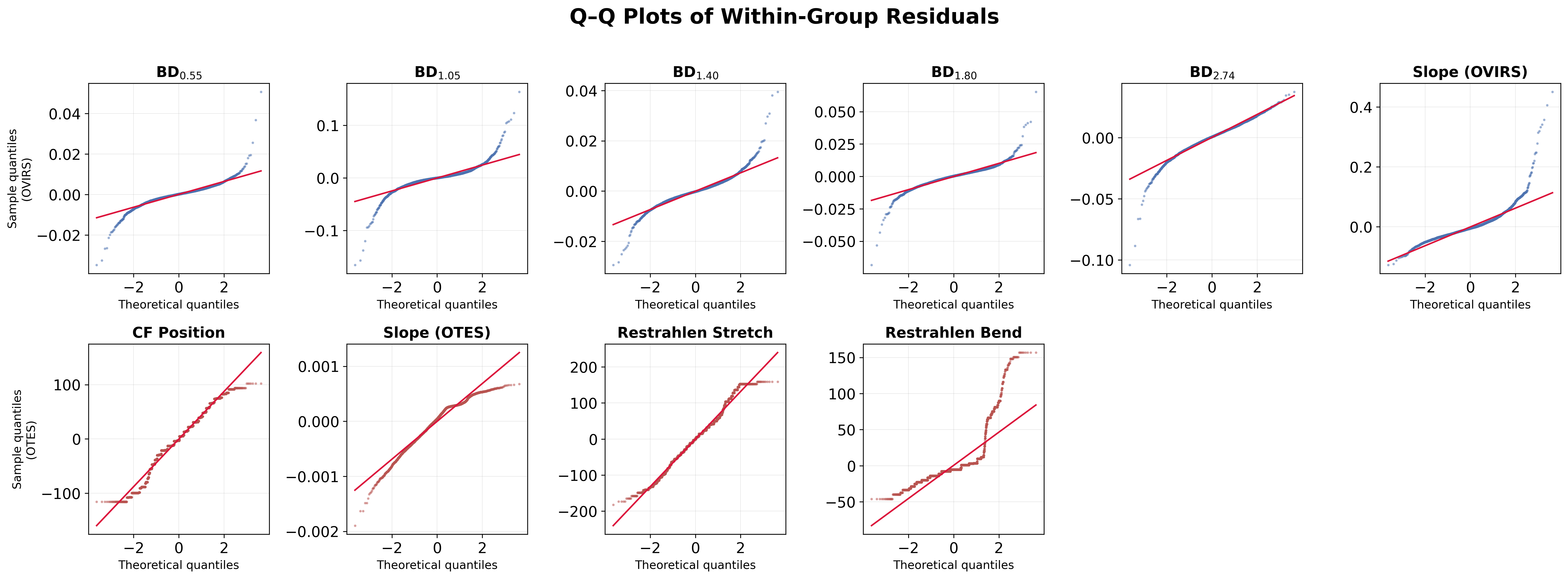}
    \caption{ANOVA Q-Q Plots}
    \label{fig:qq}
\end{figure}

Levene's test results for all band parameters are given in Table~\ref{tab:levene}; Welch's ANOVA results replacing the standard one-way ANOVA are given in Table~\ref{tab:welch_anova}. Q-Q plots of within-group residuals are shown in Figure~\ref{fig:qq}.

\begin{table}[htbp]
\centering
\caption{Levene's test for homogeneity of variance across the four
candidate sampling sites for each spectral band parameter.
Parameters with $p < 0.05$ are heteroscedastic; those were re-examined
with Welch's one-way ANOVA (Table~\ref{tab:welch_anova}).}
\label{tab:levene}
\begin{tabular}{llccc}
\hline
Instrument & Parameter & $F$ & $p$-value & Homoscedastic? \\
\hline
\multicolumn{5}{l}{\textit{OVIRS}} \\
& BD$_{0.55}$ & $659.36$ & $< 10^{-300}$ & \textbf{No} \\
& BD$_{1.05}$ & $1001.70$ & $< 10^{-300}$ & \textbf{No} \\
& BD$_{1.40}$ & $606.68$ & $< 10^{-300}$ & \textbf{No} \\
& BD$_{1.80}$ & $939.13$ & $< 10^{-300}$ & \textbf{No} \\
& BD$_{2.74}$ & $785.48$ & $< 10^{-300}$ & \textbf{No} \\
& Spectral Slope & $318.06$ & $3.12 \times 10^{-203}$ & \textbf{No} \\
\hline
\multicolumn{5}{l}{\textit{OTES}} \\
& CF Position (cm$^{-1}$) & $229.02$ & $2.40 \times 10^{-146}$ & \textbf{No} \\
& Spectral Slope & $96.24$ & $6.96 \times 10^{-62}$ & \textbf{No} \\
& Reststrahlen Stretch (cm$^{-1}$) & $26.37$ & $5.10 \times 10^{-17}$ & \textbf{No} \\
& Reststrahlen Bend (cm$^{-1}$) & $250.14$ & $1.18 \times 10^{-159}$ & \textbf{No} \\
\hline
\end{tabular}
\end{table}

\begin{table}[htbp]
\centering
\caption{Welch's one-way ANOVA for spectral band parameters that
failed Levene's test ($p < 0.05$). Degrees of freedom use the
Welch--Satterthwaite approximation.}
\label{tab:welch_anova}
\begin{tabular}{llcccc}
\hline
Instrument & Parameter & $F$ & $df_1$ & $df_2$ & $p$-value \\
\hline
\multicolumn{6}{l}{\textit{OVIRS}} \\
& BD$_{0.55}$ & $208.68$ & $3$ & $12588.7$ & $1.11 \times 10^{-16}$ \\
& BD$_{1.05}$ & $252.19$ & $3$ & $12810.3$ & $1.11 \times 10^{-16}$ \\
& BD$_{1.40}$ & $16.34$ & $3$ & $12176.4$ & $1.35 \times 10^{-10}$ \\
& BD$_{1.80}$ & $573.85$ & $3$ & $12860.7$ & $1.11 \times 10^{-16}$ \\
& BD$_{2.74}$ & $727.45$ & $3$ & $11314.3$ & $1.11 \times 10^{-16}$ \\
& Spectral Slope & $509.83$ & $3$ & $11532.8$ & $1.11 \times 10^{-16}$ \\
\hline
\multicolumn{6}{l}{\textit{OTES}} \\
& CF Position (cm$^{-1}$) & $314.67$ & $3$ & $8280.5$ & $1.11 \times 10^{-16}$ \\
& Spectral Slope & $728.73$ & $3$ & $9015.9$ & $1.11 \times 10^{-16}$ \\
& Reststrahlen Stretch (cm$^{-1}$) & $339.57$ & $3$ & $8460.8$ & $1.11 \times 10^{-16}$ \\
& Reststrahlen Bend (cm$^{-1}$) & $285.99$ & $3$ & $9649.4$ & $1.11 \times 10^{-16}$ \\
\hline
\end{tabular}
\end{table}

\end{article}
\clearpage